\documentclass[%
 notitlepage
 amsmath,
 aps,
]{revtex4-2}
\usepackage{amsmath}
\usepackage{bm}
\usepackage{natbib}
\usepackage{graphicx}
\usepackage{hyperref}
\usepackage{epstopdf}
\usepackage{color,soul}

\newcommand{\hla}[2][white]{{\sethlcolor{#1}\hl{#2}}}
\newcommand{\hlb}[2][white]{{\sethlcolor{#1}\hl{#2}}}
\newcommand{\hlc}[2][white]{{\sethlcolor{#1}\hl{#2}}}

\begin{document}
\title{A mesh-free framework for high-order direct numerical simulations of combustion in complex geometries}

\author{J. R. C. King}
\email{jack.king@manchester.ac.uk}
\affiliation{School of Engineering, University of Manchester, Manchester, UK}
\date{\today}
\begin{abstract}

The multiscale nature of turbulent combustion necessitates accurate and computationally efficient methods for direct numerical simulations (DNS). The field has long been dominated by high-order finite differences, which lack the flexibility and adaptivity for simulations of complex geometries and flame-turbulence-structure interactions in realistic settings. In this work I introduce a new approach to DNS of premixed combustion, based on a high-order mesh-free discretisation in combination with finite differences, enabling high-order simulations in non-trivial geometries. The approach is validated against a range of two- and three-dimensional flows, both laminar and turbulent, and reacting and inert. The present method a) has the resolving power for DNS of both laminar flames and inert turbulence with comparable accuracy to high-order finite differences, b) can capture the dynamics of unsteady bluff body stabilised flames, and c) is capable of simulating flame-turbulence interactions, with results comparing qualitatively well with published data.
This work paves the way for DNS of combustion in complex geometries, offering an alternative approach to methods based on structured grids with immersed boundaries, or unstructured meshes. Further studies with the present method are proposed, which will aid understanding of fundamental flame dynamics in non-trivial geometries. Planned developments in adaptivity and extension of the mesh-free construction to all three dimensions will increase the value of the method, and support the push towards DNS of real geometries. 
\end{abstract}

\maketitle

\section{Introduction}\label{sec:intro}

Turbulent combustion is multiscale in nature, with chemical time scales of the order of nanoseconds, and sub-millimetre flame fronts, interacting highly non-linearly with much larger scale hydrodynamic, turbulent, and acoustic structures. Consequently, any direct numerical simulation (DNS) of combustion needs to be both highly accurate and computationally efficient. The need for high-order schemes is widely acknowledged in the literature (e.g.~\cite{emmett_2014,emmett_2019,zirwes_2020,domingo_2023}), and has led to the field being dominated by high-order structured-grid methods, in particular finite differences. For a comprehensive review of the state-of-the-art in DNS of combustion, I refer the reader to~\cite{domingo_2023}. 

As we move towards hydrogen fuels, DNS will continue to play a central role in improving our understanding of fundamental flame dynamics, providing insight impossible to obtain from experiments. Flame-turbulence-structure interactions are one key area of relevance, however current methods lack the ability to handle complex geometries. In the arena of high-order finite differences, codes such as SENGA+~\cite{cant_2012}, S3D~\cite{chen_2011} and KARFS~\cite{perez_2018} have had significant impact on combustion DNS. These are based on tenth-order (SENGA+) and eighth-order (S3D and KARFS) central finite difference schemes, and numerically solve the fully compressible reacting Navier-Stokes equations using explicit Runge-Kutta schemes for time advancement. Whilst extremely fast and accurate, these codes are limited to simple geometries, and lack the flexibility to simulate complex flame-structure interactions, whilst their construction largely precludes adaptivity. \hlb{Body-fitted or block-structured meshes can be implemented for finite difference codes, but only in geometries of limited complexity.} Another relevant structured-mesh code is the finite volume code HAMISH~\cite{cant_2022}, which uses unstructured refinement on a Cartesian grid, along with fourth-order flux reconstruction to give an adaptive code with accuracy levels sufficient for DNS of turbulent combustion, though this remains limited at present to simple geometries. Complex geometries may be included in structured-mesh codes via immersed \hlb{or embedded} boundary methods, and in recent years several leading codes have benefitted from development in this area (e.g. S3D~\cite{rauch_2018}), although \hlb{these immersed boundary methods} are generally limited to second order accuracy in practice\hlb{. Simulations of turbulent combustion involve very fine flame fronts which dynamically evolve, but generally occupy only a small portion of the domain. This is a setting where adaptivity can bring significant benefits in reduced computational expense, and adaptivity is a highly desirable property for the next generation of combustion DNS codes~\mbox{\cite{cant_2022}}. However, finite difference codes face challenges in this regard: variable resolution is limited to grid or coordinate stretching techniques, and adaptivity is generally not possible~\mbox{\cite{verzicco_2023}}. Despite the advantages in accuracy and efficiency, the need for the next generation of combustion codes to include adaptivity and complex geometry capabilities has resulted in a move towards more flexible alternatives}.

Unstructured-mesh methods provide an alternative with increased geometric flexibility, as a body fitted mesh may \hlb{more easily} be constructed to accurately represent the desired geometry. An example of a leading unstructured-mesh code in combustion DNS is AVBP, which uses the cell-vertex finite volume method, and although it is limited to low order spatial discretisation, it has been widely employed for simulations of combustion in non-trivial geometries (e.g.~\cite{migbreb_2016}). Codes based on the spectral difference method~\cite{marchal_2023}, in which high-order polynomial reconstruction of both solution and fluxes in each grid-element result in a highly accurate method on an unstructured mesh, show significant promise for the future of combustion DNS~\cite{domingo_2023}.

Where acoustics are not of interest, the variable density low-Mach approximation allows simulations with a larger time-step, potentially providing computational speedup. In the structured-mesh field, a key example of this approach is the adaptive-mesh code PELE~\cite{nonaka_2018} (and its subvariants), which solves the variable-density low-Mach multispecies Navier-Stokes equations with a second-order scheme, with recently developed embedded boundary capabilities~\cite{sitaraman_2021}. The elimination of the acoustic time-step constraint in the low-Mach approximation makes weak-form methods (which require solution of large, sparse linear systems every time-step) feasible. In this area, the leading Spectral Element method (SEM) code Nek5000~\cite{nek5000} has been adapted to low-Mach DNS of combustion in complex geometries (e.g.~\cite{schmitt_2016}). The crucial advantage of the SEM over finite difference and finite volume schemes is that it delivers extremely high accuracy on unstructured meshes and in complex geometries. However, where acoustics are of interest (as in the present case, and many practical combustion applications), weak-form methods such as SEM are prohibitively expensive. A second limitation, which applies to SEM and all unstructured-mesh methods, is the challenge of constructing a high-quality body-fitted mesh. In complex geometries the construction of a high-quality mesh, where there is no significant loss of accuracy due to skewed mesh elements, can be an extremely resource-intensive task, and in some cases (e.g. porous media), can take longer than the simulation itself~\cite{wood_2020}. This leads us to consider a fundamentally different approach: mesh-free methods.

Mesh-free methods are an alternative to grid-based or mesh-based methods, with Smoothed Particle Hydrodynamics (SPH) probably the most widely used (see~\cite{lind_2020} for a recent review). In the mesh-free literature, computational points are usually referred to as particles: an intuitive description as mesh-free methods are often constructed in a Lagrangian framework, where the collocation points move with the fluid. However, even in an Eulerian framework, mesh-free methods can have advantages over mesh-based methods, particularly for simulations on complex geometries. In mesh-based methods, the discretisation requires both the positions of computational points, and their topological connectivity. In mesh-free methods, we rely only on the relative positions of computational points. With no information on node connectivity required, discretising a geometry with an unstructured node-set (for example via propagating front algorithms~\cite{fornberg_2015a}) is straightforward and easily automated. A limitation of many mesh-free methods is accuracy: for those in a Lagrangian frame, the construction of high-order interpolants at every time-step is prohibitively expensive, and hence the interpolation used to construct derivative approximations is typically low-order. For example, the standard SPH discretisation is formally zeroth order, or even divergent~\cite{quinlan}, and the resolving power falls far short of that required for DNS at reasonable resolutions~\cite{mayrhofer_2015}. Even many of the proposed consistency corrections are only first, second or at most third order (e.g.~\cite{bonet_lok,zhang_2004,asprone_2010,asprone_2011,sibilla_2015,liu_1995,benito_2007,gavete_2017}).

With these limitations in mind, I recently developed the Local Anisotropic Basis Function method (LABFM)~\cite{king_2020}, and demonstrated its potential with simulations of inert isothermal flows in~\cite{king_2022}. LABFM is a method for obtaining finite-difference style weights on a stencil of arbitrarily distributed unstructured nodes, with no information on node connectivity required. The weights are obtained by solution of local linear systems constructed to ensure that consistency is satisfied up to the desired order, with consistency between $4^{th}$ and $10^{th}$ order demonstrated~\cite{king_2022}. \hla{In~\mbox{\cite{king_2020}} we assessed the eigenvalues of operators constructed with LABFM, which show very low levels of numerical dissipation, whilst i}n~\cite{king_2022}, we showed that the resolving power of LABFM is comparable, and in some cases outperforms, classical central finite differences\hla{ of equivalent order}. Compared with other consistent mesh-free methods (for a review of which, I refer the reader to the introductions of~\cite{king_2020,king_2022}), LABFM may be formulated at higher order, and has a lower computational cost for a given level of accuracy. LABFM has significant potential in the field of combustion DNS, with accuracy and stability properties akin to high-order finite differences, and excellent scalability, whilst retaining the geometric flexibility common to mesh-free methods.

In this work I demonstrate the potential for mesh-free DNS of combustion in non-trivial geometries. With LABFM and high-order finite differences having a similar construction, I show how they can be combined for three-dimensional simulations where the geometry is homogeneous in one dimension (e.g. flow past a cylinder). With this combination of LABFM and high-order finite differences, I vary the order of the spatial discretisation is between $4^{th}$ and $10^{th}$, whilst using an explicit $3^{rd}$ Runge-Kutta scheme for time integration. The method is validated against a range of canonical inert and reacting flow problems, showing the framework has the resolving power required for both combustion and turbulence DNS.

The remainder of the paper is set out as follows. In Section~\ref{ge} I introduce the governing equations. In Section~\ref{ni} I describe the numerical implementation of the governing equations within the present framework, including details the LABFM/FD based discretisation, filtering procedure, and boundary conditions. Section~\ref{nt} contains the results of numerical tests on two- and three-dimensional reacting and turbulent flows. \hlc{In Section~\mbox{\ref{sec:cp}} I discuss the computational performance and parallel scaling of the method.} Section~\ref{conc} is a summary of conclusions.

Before proceding further, I make a brief comment on my notation. The letter $m$ is used to denote the polynomial consistency of the numerical scheme. Subscripts $\alpha$ and $\beta$ are used to denote chemical species. Of the Latin subscripts, $i$, $j$, and $k$ are reserved for vector indices in Einstein notation (and where these indices are repeated, summation is implied), whilst $a$ and $b$ are reserved for node (collocation point) indices. All other symbols are defined at the point of first use.

\section{Governing Equations}\label{ge}

I consider the compressible Navier-Stokes equations in conservative form for a mixture of $N_{S}$ miscible reacting chemical species
\begin{subequations}\begin{align}
\frac{\partial\rho}{\partial{t}}+\frac{\partial\rho{u}_{k}}{\partial{x}_{k}}=&~0\label{eq:mass}\\
\frac{\partial\rho{u}_{i}}{\partial{t}}+\frac{\partial\rho{u}_{i}u_{k}}{\partial{x}_{k}}=&-\frac{\partial{p}}{\partial{x}_{i}}+\frac{\partial\tau_{ki}}{\partial{x}_{k}}+\rho{f}_{i}\label{eq:mom}\\
\frac{\partial\rho{E}}{\partial{t}}+\frac{\partial\rho{u}_{k}E}{\partial{x}_{k}}=&-\frac{\partial{p}u_{k}}{\partial{x}_{k}}-\frac{\partial{q}_{k}}{\partial{x}_{k}}+\frac{\partial\tau_{ki}u_{i}}{\partial{x}_{k}}+\rho\displaystyle\sum_{\alpha=1}^{N_{S}}Y_{\alpha}f_{k}\left({u}_{k}-V_{\alpha,k}\right)\label{eq:en}\\
\frac{\partial\rho{Y}_{\alpha}}{\partial{t}}+\frac{\partial\rho{u}_{k}Y_{\alpha}}{\partial{x}_{k}}=&~\dot\omega_{\alpha}-\frac{\partial\rho{V}_{\alpha,k}Y_{\alpha}}{\partial{x}_{k}}\qquad\forall\alpha\in\left[1,N_{S}\right]\label{eq:Y}
\end{align}\label{eq:ge}
\end{subequations}
where $\rho$ is the density, ${u}_{i}$ the $i$-th component of velocity, $p$ is the pressure, $E$ is the total energy, $\tau_{ki}$ are components of the viscous stress tensor, $f_{i}$ is the $i$-th component of body force, $q_{k}$ the $k$-th component of the heat flux vector, $Y_{\alpha}$ is the mass fraction of species $\alpha\in\left[1,N_{S}\right]$, and $V_{\alpha,k}$ is the $k$-th component of the molecular diffusion velocity of species $\alpha$. $\dot\omega_{\alpha}$ is the production rate of species $\alpha$. The total energy is related to the other thermodynamic quantities by
\begin{equation}E=\displaystyle\sum_{\alpha}h_{\alpha}Y_{\alpha}-\frac{p}{\rho}+\frac{1}{2}{u}_{i}{u}_{i},\label{eq:E}\end{equation}
where $h_{\alpha}$ is the enthalpy of species $\alpha$. The system is closed with an equation of state
\begin{equation}\frac{p}{\rho}={R}_{mix}T=\frac{R_{0}T}{W_{mix}}={R}_{0}T\displaystyle\sum_{\alpha}\frac{Y_{\alpha}}{W_{\alpha}}\label{eq:eos}\end{equation}
where $R_{mix}$ is the gas constant for the mixture, $R_{0}$ is the universal gas constant, $W_{\alpha}$ is the molar mass of species $\alpha$, and $W_{mix}$ is the molar mass of the mixture. In~\eqref{eq:E} and~\eqref{eq:eos}, and hereafter, sums over $\alpha$ are taken to be over all species $\alpha\in\left[1,N_{S}\right]$. The viscous stress is defined as 
\begin{equation}\tau_{ki}=\mu\left(\frac{\partial{u}_{k}}{\partial{x}_{i}}+\frac{\partial{u}_{i}}{\partial{x}_{k}}-\frac{2}{3}\frac{\partial{u}_{j}}{\partial{x}_{j}}\delta_{ik}\right)\label{eq:tau},\end{equation}
where $\mu$ is the dynamic viscosity and $\delta_{ik}$ is the Kronecker delta. The molecular diffusion term in~\eqref{eq:Y} is modelled as a Fickian process with
\begin{equation}-\rho{V}_{\alpha,k}Y_{\alpha}=\rho\hat{V}_{\alpha,k}Y_{\alpha}-Y_{\alpha}\displaystyle\sum_{\beta\in\left[1,N_{S}\right]}\rho\hat{V}_{\beta,k}Y_{\beta}\label{eq:fickdiff},\end{equation}
in which $\rho\hat{V}_{\alpha,k}Y_{\alpha}$ is obtained from the diffusion model (either constant Lewis numbers or mixtured-averaged, details below), and the final term is a correction to ensure that the compatibility condition
\begin{equation}\displaystyle\sum_{\alpha}Y_{\alpha}=1\label{eq:sumY}\end{equation}
is satisfied. The heat flux vector is given by
\begin{equation}q_{k}=-\lambda\frac{\partial{T}}{\partial{x}_{k}}+\displaystyle\sum_{\alpha}\rho{V}_{\alpha,k}Y_{\alpha}h_{\alpha}\label{eq:hfv},\end{equation}
where $\lambda$ is the thermal conductivity of the mixture.

The temperature dependencies of thermodynamic quantities (heat capacity $c_{p,\alpha}$,  enthalpy $h_{\alpha}$) for each species take polynomial form fitting the standard NASA polynomials~\cite{gordon_1971}. The energy in~\eqref{eq:E} can be expressed as a non-linear function of $T$ and the conserved variables, which is solved using a Newton-Raphson method to obtain $T$. In the specific case where $c_{p,\alpha}$ is independent of temperature, the expression for temperature becomes linear and the Newton-Raphson method converges exactly in a single iteration. I take the last known value of $T$ as a suitable guess to initialise the Newton-Raphson method, and for typical flows (e.g. premixed $H_{2}$-air flames), this results in convergence to a tolerance of $10^{-10}$ within two or three iterations.

I consider two models for transport properties. The constant Lewis number approximation, and a mixture-averaged transport model. For the case of constant Lewis numbers, I model the diffusive flux as \begin{equation}\rho\hat{V}_{\alpha,k}Y_{\alpha}=-\rho{D}_{\alpha}\frac{\partial{Y}_{\alpha}}{\partial{x}_{k}},\end{equation}
in which $D_{\alpha}$ is the molecular diffusivity of species $\alpha$. I prescribe a temperature dependence of the viscosity of the form
\begin{equation}\mu=\mu_{ref}\left(\frac{T}{T_{ref}}\right)^{r_{T}},\label{eq:tdtp_mu}\end{equation}
with $T_{ref}$ a reference temperature and $r_{T}$ a constant exponent typically set to $r_{T}=0.7$, except where otherwise specified. The thermal conductivity $\lambda$ and molecular diffusivities $D_{\alpha}$ follow from the definitions of Prandtl $Pr$ and Lewis $Le_{\alpha}$ numbers.

For the case of mixture-averaged transport, I model the diffusive flux by
\begin{equation}\rho\hat{V}_{\alpha,k}Y_{\alpha}=-\rho{D}_{\alpha}\frac{\partial{Y}_{\alpha}}{\partial{x}_{k}}-\rho{D}_{\alpha}Y_{\alpha}\left(\frac{\partial\ln{T}}{\partial{x}_{k}}+\frac{\partial\ln\rho}{\partial{x}_{k}}-\frac{W_{\alpha}}{\rho{R}_{0}T}\frac{\partial{p}}{\partial{x}_{k}}\right).\end{equation}
In this case, the transport properties and binary diffusion coefficients for each species are evaluated using polynomials (polynomial in $\ln{T}$) following~\cite{kee_1986}, and the mixture-averaged transport properties are obtained from combination rules following~\cite{ern_1994,ern_1995} with the Hirschfelder-Curtiss approximation for diffusion~\cite{hirschfelder_curtiss}. Soret and Dufour effects are neglected.

I consider a general reaction mechanism consisting of $M$ steps, written in the form
\begin{equation}\displaystyle\sum_{\alpha}\nu_{\alpha,l}^{\prime}\mathcal{M}_{\alpha}\rightarrow\displaystyle\sum_{\alpha}\nu_{\alpha,l}^{\prime\prime}\mathcal{M}_{\alpha}\qquad\forall{l}\in\left[1,M\right],\end{equation}
with the inclusion of third bodies where appropriate. Here $\mathcal{M}_{\alpha}$ symbolises the chemical symbol of species $\alpha$, and $\nu_{\alpha,l}^{\prime}$ and $\nu_{\alpha,l}^{\prime\prime}$ are the molar stoichiometric coefficients of species $\alpha$ in reaction $l$. The reaction rate $\omega_{\alpha}$ of species $\alpha$ is given by the sum of the molar production rate over all reactions. The system of equations which describes the evaluation of molar production rates is not unique to this work, wherein it closely resembles the reaction frameworks used in other codes (e.g.~\cite{babkovskaia_2011,cant_2012,nonaka_2018,cant_2022}). I omit details for brevity, but note that third bodies, backwards reaction rates and Lindemann forms are included where required. For more details I refer the interested reader to~\cite{poinsot_2005}. Although I have tested my method on a range of reaction mechanisms, in this work I focus only on two: a single-step Arrhenius mechanism designed to match the flame speed of stoichiometric methane-air combustion, and the $21$ step, $9$ species hydrogen-air mechanism of~\cite{li_2004}. For hydrogen combustion, I use the mixture-averaged transport model, whilst for the single-step methane mechanism, I assume constant Lewis numbers.

\section{Numerical implementation}\label{ni}

The spatial discretisation is based on the Local Anisotropic Basis Function Method (LABFM) in the first two dimensions, and high-order finite differences in the third dimension. This allows us to simulate complex geometries, although I am limited to those in which the geometry is homogeneous in the third dimension. I could implement a fully mesh-free formulation, with LABFM used for derivatives in all three dimensions, at the cost of increased stencil sizes. This is an area of development for the author. LABFM has been detailed and extensively analysed in~\cite{king_2020,king_2022}, and I refer the reader to these works for a complete description, where the congergence properties of the method are comprehensively demonstrated. For convenience of exposition, I denote as $\mathcal{A}_{12}$ any plane in which $x_{3}$ is constant, and $\mathcal{A}_{12}\left(C\right)$ the specific realisation of $\mathcal{A}_{12}$ in which $x_{3}=C$.

The domain is discretised in $\mathcal{A}_{12}\left(0\right)$ with an unstructured point cloud of $N_{12}$ nodes. The discretisation in the third dimension is uniform: there are $N_{3}$ copies of the $\mathcal{A}_{12}\left(0\right)$ discretisation, equispaced along the $x_{3}$ axis. Hence, there are a total of $N=N_{12}N_{3}$ nodes in the domain. Each node $a\in\left[1,N\right]$ has position $x_{a,i}$, a distribution lengthscale $s_{a}$, and a computational stencil lengthscale $h_{a}$. $s_{a}$ is the average node spacing in $\mathcal{A}_{12}\left(x_{a,3}\right)$ around node $a$, and is analagous to the grid spacing in a finite difference scheme. The spacing in $x_{3}$ is denoted $s_{3}$, and is uniform throughout the domain, whilst $s_{a}$ may vary as a function of $x_{1}$ and $x_{2}$.

\begin{figure}
\includegraphics[width=0.44\textwidth]{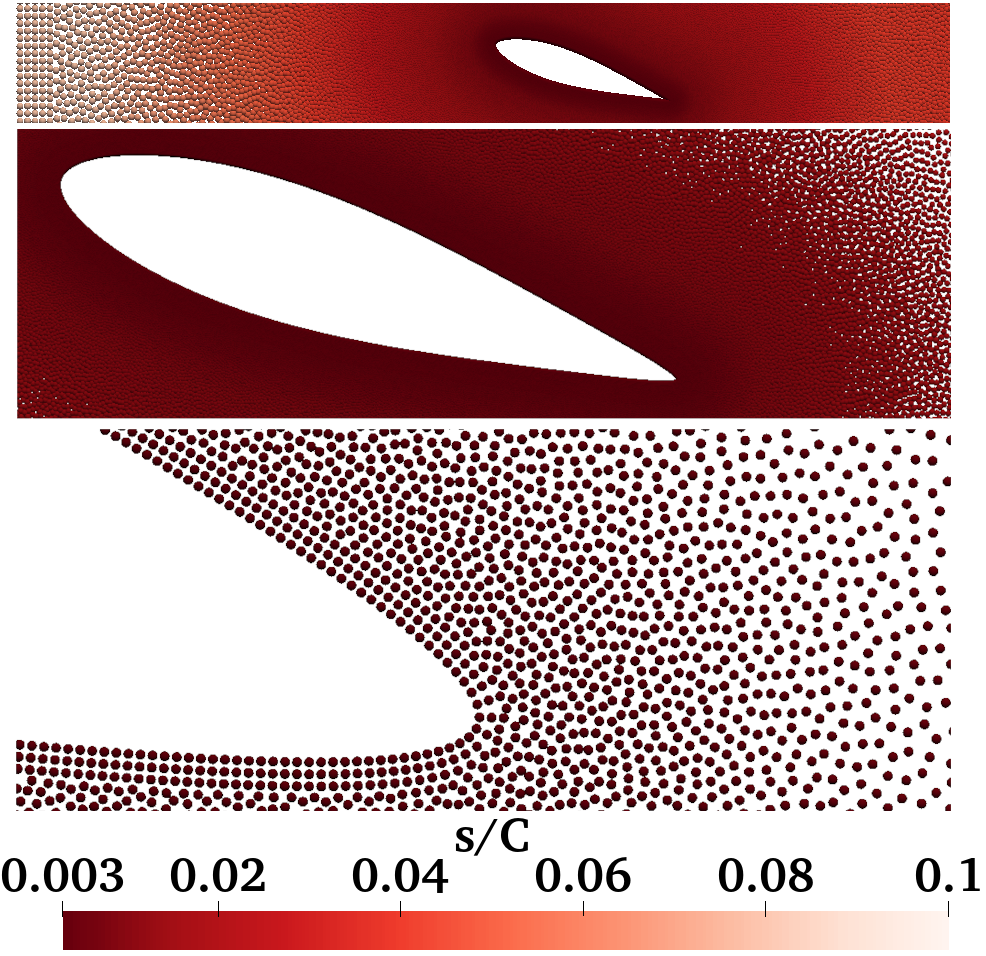}
\includegraphics[width=0.55\textwidth]{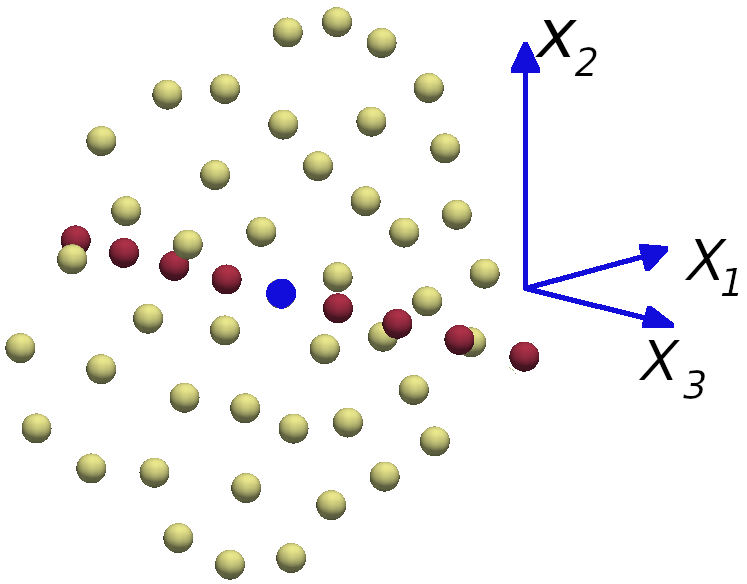}
\caption{Left panel: A typical node distribution in the $x_{1}-x_{2}$ plane $\mathcal{A}_{12}$ around an aerofoil. The color represents the local resolution lengthscale $s$ normalised by the chord length $C$. Right panel: A typical stencil for a three-dimensional simulation. The nodes shown are those in the stencil of the central node in blue. The unstructured nodes in yellow lie in the same plane $\mathcal{A}_{12}$ as the central node. The nodes in red are those in the finite difference stencil for the $x_{3}$ direction.\label{fig:disc2}}
\end{figure}

The left panel of Figure~\ref{fig:disc2} shows an example two-dimensional discretisation in the present framework for an aerofoil. Wall, inflow and outflow boundaries are discretised with a locally structured node distribution (detailed in Section~\ref{sec:bound}), whilst the remaining space is filled with an unstructured node-set. The resolution is non-uniform, and in the present example $s_{a}$ varies by a factor of approximately $30$ through the domain. The node distribution is generated using a variation of the propagating front algorithm originally developed by~\cite{fornberg_2015a}, following~\cite{king_2022}. I set $s_{3}$ to the weighted mean of $s_{a}$ taken over the domain as
\begin{equation}s_{3}=\left[\frac{1}{N}\displaystyle\sum_{a=1}^{N}s_{a}^{n_{s}}\right]^{\frac{1}{n_{s}}},\end{equation}
where $n_{s}$ takes the value $1$ for the arithmetic mean, and $2$ for the area-weighted mean, with the choice of $n_{s}$ case dependent. Unless explicitly specified, I set $n_{s}=1$ in this work. \hlb{In some cases, setting $s_{3}$ to the average of $s_{a}$ may result in the discretisation in the third dimension having insufficient resolution to capture the smallest scales of the flow (either turbulent structures or flame fronts). In all cases in the present work, a manual check sufficed to reassure that this issue did not arise. Ultimately in the present framework there is a compromise that must be made: larger $s_{3}$ may under-resolve flames and turbulent structures in some regions, whilst smaller $s_{3}$ will increase computational costs, over-resolving regions of little interest. Including adaptivity, and extending the framework to be mesh-free in all three dimensions, is a key avenue I intend to explore in future.}

Each node holds the conservative variables $\rho_{a}$, $\left(\rho{u}_{i}\right)_{a}$, $\left(\rho{E}\right)_{a}$ and $\left(\rho{Y}_{\alpha}\right)_{a}$. The governing equations are solved on the set of $N$ nodes. I denote the difference between any property $\left(\cdot\right)$ at two nodes by $\left(\cdot\right)_{ba}=\left(\cdot\right)_{b}-\left(\cdot\right)_{a}$. The computational stencil for each node $a$ is denoted $\mathcal{N}_{a}$, and is constructed to contain all nodes $b$ in the same plane $\mathcal{A}_{12}\left(x_{3,a}\right)$ such that $\lvert{x}_{i,ba}x_{i,ba}\rvert\le4h_{a}^{2}$, (with the repeated subscript $i$ implying summation), alongside all the nodes $b$ with $x_{1,ba}=x_{2,ba}=0$ required to build a finite difference stencil in the $x_{3}$ dimension. The right panel of Figure~\ref{fig:disc2} shows a typical computational stencil (for the central node in blue) for a three dimensional simulation, with the nodes in the $\mathcal{A}_{12}$ plane in yellow, and the nodes used to construct the finite difference stencil in $x_{3}$ coloured red.

\subsection{Evaluation of spatial derivatives}\label{sec:der}

In present framework, all spatial derivative operators take the form
\begin{equation}L^{d}_{a}\left(\cdot\right)=\displaystyle\sum_{b\in\mathcal{N}_{a}}\left(\cdot\right)_{ba}w^{d}_{ba},\label{eq:general_do}\end{equation}
where $d$ identifies the derivative of interest (e.g. $d=x_{1}$ indicates the operator $L^{d}$ approximates $\partial\left(\cdot\right)/\partial{x}_{1}$, and $d=x_{i}x_{i}$ (summation implied) indicates the operator approximates the Laplacian), and $w^{d}_{ba}$ are a set of inter-node weights for the operator. Note that the construction of~\eqref{eq:general_do} takes the same form for both LABFM-based and finite difference-based derivative approximations. For derivatives in the $x_{3}$ direction, the weights $w^{d}_{ba}$ are simply central finite difference weights of order $m$. For derivatives in $\mathcal{A}_{12}$, LABFM is used to set $w^{d}_{ba}$, which is constructed from a weighted sum of anisotropic basis functions centred on node $a$. Note, derivatives at boundaries require a slightly different treatment, detailed in Section~\ref{sec:bound} and Appendix~\ref{ap:ifd}. The basis functions used herein are formed from bi-variate Hermite polynomials multiplied by a radial basis function, and the basis function weights are set such that the operator achieves a specified polynomial consistency $m$. For polynomial consistency $m$, the errors in the approximation of first derivatives are $\mathcal{O}\left(s^{m}\right)$, whilst the errors in the approximation of second derivatives are $\mathcal{O}\left(s^{m-1}\right)$. Calculation of these weights is done as a pre-processing step, and described in detail in~\cite{king_2022}. 

The order of consistency of the scheme is specified between $m=4$ and $m=10$, and although there is capability for this to be spatially (or temporally) varying, in this work I set $m=8$ uniformly away from boundaries (except where explicitly stated in my investigations on the effects of changing $m$). This provides an appropriate compromise between accuracy and computational cost. At non-periodic boundaries, the consistency of the LABFM reconstruction is smoothly reduced to $m=4$, as detailed in Section~\ref{sec:bound}. The stencil scale $h_{a}$ is initialised to $h_{a}=2.7s_{a}$ in the bulk of the domain, and $h_{a}=2.4s_{a}$ near boundaries (choices \hlb{based on experience from previous studies} as being large enough to ensure stability~\cite{king_2020}). In bulk of the domain, the stencil scale is then reduced \hlb{(effectively reducing the support radius)} following the optimisation procedure described in~\cite{king_2022}. This has the effect of both reducing computational costs, and increasing the resolving power of LABFM, and $h_{a}$ subsequently takes a value slightly larger than the smallest value for which the discretisation remains stable \hlb{with the non-uniformity in $h_{a}/s_{a}$ accounting for the differences in local node distribution within each stencil}.

The evaluation of derivatives using~\eqref{eq:general_do} constitutes the bulk of the computational cost, and hence it is desirable to use analytic expressions to relate derivatives of secondary properties to those evaluated via~\eqref{eq:general_do} where possible. First derivatives of $\rho$, ${u}_{i}$, $\rho{E}$ and ${Y}_{\alpha}$, alongside the temperature $T$ and pressure $p$, are evaluated directly using~\eqref{eq:general_do}. Convective terms are then constructed from combinations thereof: for example
\begin{equation}\frac{\partial\rho{u}_{i}E}{\partial{x}_{i}}=u_{i}\frac{\partial\rho{E}}{\partial{x}_{i}}+\rho{E}\frac{\partial{u}_{i}}{\partial{x}_{i}}.\end{equation}
Laplacians and direct second (i.e. $\partial^{2}\left(\cdot\right)/\partial{x}_{i}\partial{x}_{i}$) derivatives are also evaluated with~\eqref{eq:general_do}. To avoid the explicit evaluation of cross-second derivatives of velocity, the viscous stress divergence is evaluated as
\begin{equation}\frac{\partial\tau_{ki}}{\partial{x}_{k}}=\frac{1}{\mu}\frac{\partial\mu}{\partial{x}_{k}}\tau_{ki}+\mu\left(\frac{\partial^{2}{u}_{i}}{\partial{x}_{k}\partial{x}_{k}}+\frac{1}{3}\frac{\partial^{2}{u}_{k}}{\partial{x}_{i}\partial{x}_{k}}\right)\label{eq:divtau},\end{equation}
where $\partial^{2}u_{k}/\partial{x_{i}}\partial{x_{k}}$ is the gradient of the velocity divergence, which is calculated through two iterations of first derivative evalations with~\eqref{eq:general_do}. The benefit of this approach is that it avoids the need to explicitly evaluate the derivatives $\partial^{2}/\partial{x}_{1}\partial{x}_{3}$ and $\partial^{2}/\partial{x}_{2}\partial{x}_{3}$. This allows for a smaller stencil size, and provides a significant reduction in the complexity associated with parallelising the scheme. Gradients of secondary properties are evaluated from analytic expressions in terms of gradients $T$ and $Y_{\alpha}$. The gradient of the enthalpy of species $\alpha$ is given by
\begin{equation}\frac{\partial{h}_{\alpha}}{\partial{x}_{k}}=\frac{dh_{\alpha}}{dT}\frac{\partial{T}}{\partial{x}_{k}}=c_{p,\alpha}\frac{\partial{T}}{\partial{x}_{k}},\end{equation}
and the gradient of the mixture specific heat capacity is
\begin{equation}\frac{\partial{c}_{p}}{\partial{x}_{k}}=\displaystyle\sum_{\alpha}\left[Y_{\alpha}\frac{dc_{p,\alpha}}{dT}\frac{\partial{T}}{\partial{x}_{k}}+c_{p,\alpha}\frac{\partial{Y}_{\alpha}}{\partial{x}_{k}}\right].\end{equation}
When constant Lewis numbers are assumed, additional relations are available to evaluate the spatial derivatives of transport properties:
\begin{subequations}\begin{align}
\frac{\partial\mu}{\partial{x}_{k}}&=\frac{\mu{r}_{T}}{T}\frac{\partial{T}}{\partial{x}_{k}}\\
\frac{\partial\lambda}{\partial{x}_{k}}&=\frac{\lambda{r}_{T}}{T}\frac{\partial{T}}{\partial{x}_{k}}+\frac{\lambda}{c_{p}}\frac{\partial{c_{p}}}{\partial{x}_{k}}\\
\frac{\partial\rho{D}_{\alpha}}{\partial{x}_{k}}&=\frac{D_{\alpha}r_{T}}{T}\frac{\partial{T}}{\partial{x}_{k}}.\end{align}\end{subequations}
For the mixture averaged model, gradients of $\mu$, $\lambda$ and $\rho{D}_{\alpha}$ are obtained directly using~\eqref{eq:general_do}. Even with these substitutions, for three-dimensional simulations the constant Lewis number transport model requires $3\left(8+N_{S}\right)$ first derivative evaluations, and $4+N_{S}$ Laplacian evaluations to construct the RHS of the governing equations. For the mixture-averaged transport model, this increases to $3\left(10+2N_{S}\right)$ gradients evaluations and $6+N_{S}$ Laplacian evaluations. Hence any reduction in stencil size whilst retaining accuracy and numerical stability has the potential to significantly reduce computational costs, and the optimisation of stencils is an active area of research for the author.

\subsection{Boundaries}\label{sec:bound}

The boundary framework extends the approach described for isothermal flows in~\cite{king_2022} in two ways. Firstly, it is extended to accommodate thermal and reacting flows, utilising the formulation of~\cite{sutherland_2003}. Secondly, for wall boundaries, a LABFM-based interpolation is used to partially replace the finite difference stencil, allowing for the band of uniformly distributed nodes along boundaries to be reduced from $5$ rows to $3$, permitting greater geometric flexibility whilst retaining the $4^{th}$ order consistency of the derivative operators at the boundaries. 

\begin{figure}
\includegraphics[width=0.99\textwidth]{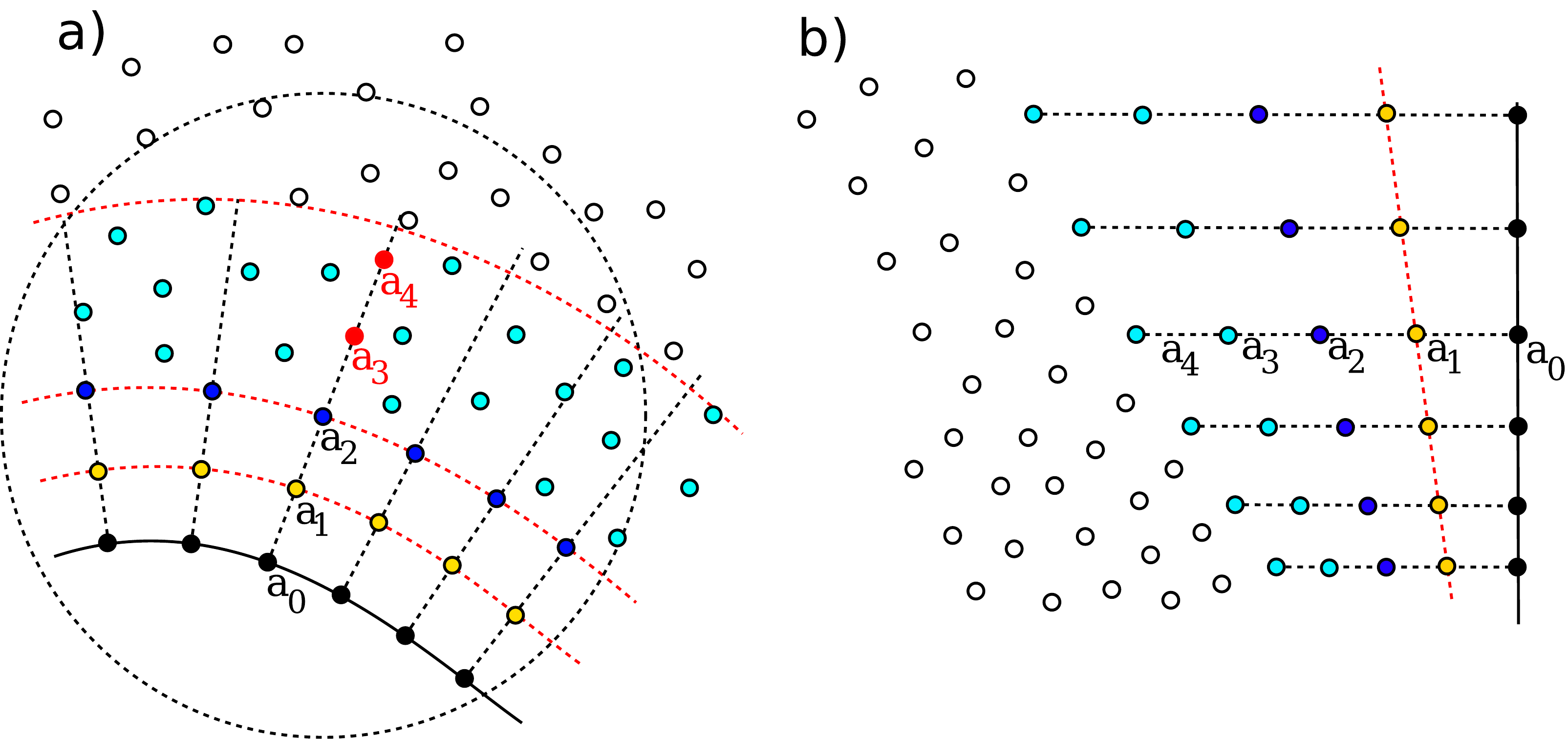}
\caption{A schematic of the discretisation near boundaries, for a) walls, and b) inflows and outflows. The solid black lines indicate a boundary, and the small circles indicate computational nodes. The straight dashed black lines indicate the boundary normal direction. For walls, the fictitious nodes required to complete the finite difference stencil are shown in red. Near-boundary nodes, at which I impose the limit $m=6$ are shown in light blue. Internal nodes are shown with a white fill. For wall boundaries, the red dashed lines indicate the regions at which different orders are imposed, and the dashed black circle shows the computational stencil used for the fictitious node interpolation.\label{fig:interpfd}}
\end{figure}

The following discretisation procedure applies to derivatives in the plane $\mathcal{A}_{12}$ only. All derivatives in the third dimension use high-order finite differences. Panel a) of Figure~\ref{fig:interpfd} illustrates the node distribution and computational stencil used near wall boundaries. Nodes are placed along the boundary (black nodes). For every boundary node $a_{0}$, and additional node $a_{1}$ (yellow, in the left panel of Figure~\ref{fig:interpfd}) is placed with position ${x}_{a_{1},i}={x}_{a_{0},i}+s_{a_{0}}{n}_{a_{0},i}$ for $i=1,2,3$, where ${n}_{a_{0},i}$ is the $i$-th component of the unit vector normal to the boundary directed towards the fluid, and a node $a_{2}$ (shown in dark blue) is placed with position ${x}_{a_{1},i}={x}_{a_{0},i}+2s_{a_{0}}{n}_{a_{0},i}$. For inflow and outflow boundaries, I place a further two nodes (denoted $a_{3}$ and $a_{4}$) along the normal vector to complete the $5$ point finite difference stencil. This formulation for inflow and outflow boundaries is as detailed in~\cite{king_2022}, and is illustrated in Figure~\ref{fig:interpfd} b). For wall boundaries, I don't complete the finite difference stencil, instead discretising the remainder of the fluid with an unstructured node set (nodes coloured light blue near the boundary, and white a distance of more than $4.5s_{a}$ from the boundary, in panel a) of Figure~\ref{fig:interpfd}). To construct the finite difference approximation to derivatives normal to the wall boundary on nodes $a_{0}$ and $a_{1}$, I interpolate fluid properties onto a set of two fictitious nodes (again denoted $a_{3}$ and $a_{4}$, coloured red in panel a) of Figure~\ref{fig:interpfd}) to complete the stencil. For clarity, the governing equations are not evolved on these fictitious nodes - these nodes are only used for interpolation to construct the finite difference stencil. \hlb{Properties are interpolated from regular nodes onto the fictitious nodes $a_{3}$ and $a_{4}$.} I refer to this scheme as an interpolated finite difference scheme. Full details are given in Appendix~\ref{ap:ifd}. 

Details of the discretisation scheme used for different nodes near the boundary are provided in Table~\ref{tab:boundaries}, where the distance between a node and the boundary is denoted $d_{a}$. Beyond the structured nodes $a_{0}$ and $a_{1}$, I use LABFM for all derivatives in the plane $\mathcal{A}_{12}$. For nodes $a_{2}$ I set $m=4$, and for nodes with $2<d_{a}/s_{a}\le4.5$ I impose the limit $m\le6$. For derivatives tangential to the boundary at nodes $a_{0}$ and $a_{1}$, I use the one-dimensional form of LABFM described in~\cite{king_2022}, to which I refer the reader for \hlb{complete} details. \hlb{For nodes $a_{0}$, the stencil for tangential derivatives involves only other nodes on the boundary (i.e. the solid black nodes in Figure~\mbox{\ref{fig:interpfd}}), and for nodes $a_{1}$, the stencil contains only nodes a distance $s_{a}$ from the boundary (i.e. the yellow nodes in Figure~\mbox{\ref{fig:interpfd}}). With these stencils, difference operators are constructed using a one-dimensional formulation of LABFM as detailed in the appendix of~\mbox{\cite{king_2022}}}. On boundaries, derivative operators are constructed in the boundary frame of reference, whilst in the first row, a rotation is applied to transform them into the Cartesian frame, as described in~\cite{king_2022}. Where the resolution is non-uniform along the boundary (as illustrated in panel b) of Figure~\ref{fig:interpfd}), an additional rotation is applied to the first derivatives evaluated on node $a_{1}$.

In terms of the numerical boundary conditions, no-slip walls (either isothermal or adiabatic), inflows (either non-reflecting or ``hard'' with specified $u_{i}$ and $T$), and non-reflecting outflows are imposed through the Navier Stokes characteristic boundary condition (NSCBC) formalism (see, e.g.~\cite{poinsot_2005}). For non-reflecting inflows, I include the transverse relaxation terms following~\cite{yoo_2007}. \hlb{As is standard in the NSCBC formalism, non-reflecting outflows are constructed to track a desired pressure $p_{out}$, introducing a small degree of acoustic reflection in return for ensuring the outflow pressure does not drift over time.} For outflows, I follow the formulation of~\cite{sutherland_2003}. The improved outflow conditions of~\cite{yoo_2007} have been implemented, but were found to be less stable than the method of~\cite{sutherland_2003}, particularly when handling passage of flames at angles oblique to the boundary. Further investigations are required in this area.

\begin{table}
\begin{center}
\caption{Schemes for derivatives near wall boundaries. IFD indicates the interpolated finite difference scheme described in detail in Appendix~\ref{ap:ifd}.\label{tab:boundaries}}
\begin{tabular}{|l|c|c|c|c|c||}
\hline
\textbf{Node} &Distance from boundary & First derivatives & Second derivatives & Filtering operator \\
\hline
$a_{0}$ & $d_{a}=0$ & IFD + 1D-LABFM & IFD + 1D-LABFM & IFD + 1D-LABFM \\
$a_{1}$ &$d_{a}=s_{a}$ & IFD + 1D-LABFM & LABFM ($m=4$) & LABFM ($m=4$) \\
$a_{2}$ & $d_{a}/2s_{a}$ & LABFM ($m=4$) & LABFM ($m=4$) & LABFM ($m=4$)  \\
-- & $2<d_{a}/s_{a}\le4.5$ & LABFM ($m\le6$) & LABFM ($m\le6$) & LABFM ($m\le6$) \\
-- & $d_{a}/s_{a}>4.5$ & LABFM ($m\le10$) & LABFM ($m\le10$) & LABFM ($m\le10$) \\
\hline
\end{tabular}
\end{center}
\end{table}

\subsection{Temporal integration and filtering}

With the right hand sides of~\eqref{eq:ge} evaluated using the derivative operators described in Section~\ref{sec:der}, and chemical source terms evaluated explicitly, the governing equations are integrated in time using an explicit Runge-Kutta scheme. I use a four-step, third-order, low-storage scheme, with embedded second order error estimation, denoted RK3(2)4[2R+]C in the classification system of~\citet{kennedy_2000}. The value of the time step is set adaptively using a PID controller to keep time-integration errors below a specified threshold (set to $10^{-4}$ in throughout the present work). 
As is common in high-order collocated methods, the low levels of numerical dissipation in the spatial derivative operators result in a scheme in which high-wavenumber modes of instability can develop. To avoid this, the solution is de-aliased after each time-step by high-order spatial filtering using the filters described in~\cite{king_2022} combined with a finite difference filter following~\cite{kennedy_1994}. The combined filter is defined for even $m$ by
\begin{equation}F\left(\phi\right)=\left\{1-\left(-1\right)^{m/2}\left(\frac{s_{3}}{2}\right)^{m}\frac{\partial^{m}}{\partial{x}_{3}^{m}}\right\}\left[\left(1+\kappa_{m}\nabla_{\mathcal{A}_{12}}^{m}\right)\phi\right]\label{eq:filt},\end{equation}
where $\kappa_{m}$ is a filter coefficient, specific to each node, calculated as in~\cite{king_2022}, and $\nabla_{\mathcal{A}_{12}}^{m}$ is the two-dimensional Laplacian raised to the power of $m/2$ (as used in the filtering in~\cite{king_2022}), defined in index notation as
\begin{equation}\nabla_{\mathcal{A}_{12}}^{m}=\displaystyle\sum_{n=0}^{m}{m\choose n}S\frac{\partial^{m}}{\partial{x}_{1}^{m-n}\partial{x}_{2}^{n}}.\end{equation}
The term within the square brackets in~\eqref{eq:filt} is the two-dimensional filter in $\mathcal{A}_{12}$, whilst the operator outside the square brackets is the filter in the $x_{3}$ dimension. The filters are applied sequentially in this manner to ensure high wavenumber modes of all orientations are suppressed, rather than just those in and orthogonal to $\mathcal{A}_{12}$. This approach follows that used in finite difference codes, where the three coordinates are filtered sequentially. With the LABFM filter, the entirety of $\mathcal{A}_{12}$ is filtered in one pass, because the filter includes cross-derivative terms, and is hence able to suppress all high wavenumber modes in $\mathcal{A}_{12}$ simultaneously, regardless of their in-plane orientation. For density, momenta and energy, I simply apply the filter as defined by~\eqref{eq:filt}, such that the filtered variables passed to the next time step are
\begin{equation}\widetilde{\phi}=F\left(\phi\right)\qquad\text{for}\quad\phi=\rho,\rho{u}_{i},\rho{E}.\end{equation}
As with the molecular diffusion terms, the filtering procedure does not guarantee satisfaction of the compatibility criteria~\eqref{eq:sumY}. Hence, I introduce a correction to the filtering procedure for the species mass fractions, which takes inspiration from the diffusion correction velocity. For the species mass fractions, I modify the procedure defining 
\begin{equation}\widetilde{\rho{Y}_{\alpha}}=F\left(\rho{Y}_{\alpha}\right)+\frac{\widetilde{\rho{Y}_{\alpha}}}{\widetilde{\rho}}F_{c}\label{eq:fc5},\end{equation}
where the filter correction $F_{c}$ is given by
\begin{equation}F_{c}=\widetilde{\rho}-\displaystyle\sum_{\alpha}F\left(\rho{Y}_{\alpha}\right)\label{eq:fc2}.\end{equation}
This ensures that
\begin{equation}\displaystyle\sum_{\alpha}\widetilde{Y}_{\alpha}=1,\label{eq:filtsumY}\end{equation}
where $\widetilde{Y}_{\alpha}=\widetilde{\rho{Y}_{\alpha}}/\widetilde\rho$, is satisfied by construction. The weighting $\widetilde{\rho{Y}_{\alpha}}/\widetilde{\rho}$ ensures that the correction is distributed across the species in proportion to their filtered mass fractions. Hence, as $Y_{\alpha}$ approaches zero, the correction for species $\alpha$ also approaches zero, ensuring that a locally negative $F_{c}$ doesn't push $\widetilde{Y}_{\alpha}<0$ locally. Equation~\eqref{eq:fc5} can be straightforwardly re-arranged to give the correction in multiplicative form as
\begin{equation}\widetilde{\rho{Y}_{\alpha}}=\frac{1}{F^{\star}_{c}}F\left(\rho{Y}_{\alpha}\right)\label{eq:fc3},\end{equation}
where the correction term $F^{\star}_{c}$ is simply
\begin{equation}F^{\star}_{c}=\frac{1}{\widetilde\rho}\displaystyle\sum_{\alpha}F\left(\rho{Y}_{\alpha}\right)\label{eq:fc4}.\end{equation}
I use the correction procedure defined by~\eqref{eq:fc3} and~\eqref{eq:fc4} throughout this work, for which the error in satisfaction of~\eqref{eq:filtsumY} remains approximately $1\times{10}^{-18}$ at all times.

\subsection{Parallelisation}

The present implementation of the numerical scheme is accelerated with a shared-distributed framework, using both OpenMP and MPI. Non-uniform block-structured three-dimensional domain decomposition is used, with the physical sizes of subdomains set such that each MPI rank contains the same number of nodes. The decomposition takes as input the desired number of processors in each dimension $N_{P1}$, $N_{P2}$, and $N_{P3}$, giving a total of $N_{P}=N_{P1}N_{P2}N_{P3}$ processors. The allocation of nodes to processors then forms a preprocessing step, following the algorithm:
\begin{enumerate}
\item Generate the node discretisation in $\mathcal{A}_{12}\left(0\right)$.
\item Re-arrange the nodes in order of increasing $x_{1}$.
\item Split the discretisation into $N_{P1}$ slices each containing the same number of nodes.
\item For each slice
\begin{enumerate}
\item Re-arrange the nodes in order of increasing $x_{2}$.
\item Split the slice into $N_{P2}$ blocks each containing the same number of nodes.
\item For each block, create the $x_{3}$ discretisation by creating $N_{3}$ copies equispaced along the $x_{3}$ axis, and distribute into $N_{P3}$ processor domains.
\end{enumerate}
\end{enumerate}
This approach results in each processor containing the same number of nodes, and the variation in processor load arises only from variations in the number of neighbours $\mathcal{N}$. Providing the number of nodes in each MPI rank is large enough that the time for MPI communications is small relative to the time for all other calculations (which scale almost perfectly with increasing $N_{P}$), the overall parallel efficiency of the scheme is excellent. In strong scaling tests, with $N=2\times{10}^{6}$ nodes, I found the scheme has a parallel efficiency of $96.5\%$ for $N_{P}=1024$ (relative to a baseline of $N_{P}=4$). I note that for smaller $N/N_{P}$, the limitation on the parallel efficiency is the load balancing, rather than the cost of MPI communications. Although each MPI rank contains the same number of nodes, the size of computational stencils $\mathcal{N}$ is not uniform (in particular, $\mathcal{N}$ is larger where resolution gradients are larger). The decrease in parallel efficiency at higher $N_{P}$ is because the load balancing is less effective as $N/N_{P}$ is reduced. For typical simulations, such as those presented in Section~\ref{nt} of two-dimensional bluff body flames and three-dimensional flame-turbulence interactions, I use $N/N_{P}\approx5000$. Although the decomposition and load balancing are sufficient for the present work, the implementation of a dynamic domain decomposition framework, with on-the-fly load balancing, is a key area of interest for the author. 

\section{Numerical tests}\label{nt}

With the convergence and stability properties of LABFM extensively investigated in~\cite{king_2020}, and the ability of the method to accurately simulate two-dimensional isothermal flows established in~\cite{king_2022}, I focus here on tests which demonstrate the performance of the framework for reacting and turbulent flows. 

\subsection{Laminar hydrogen-air flames}\label{sec:1d}

\begin{figure}
\includegraphics[width=0.99\textwidth]{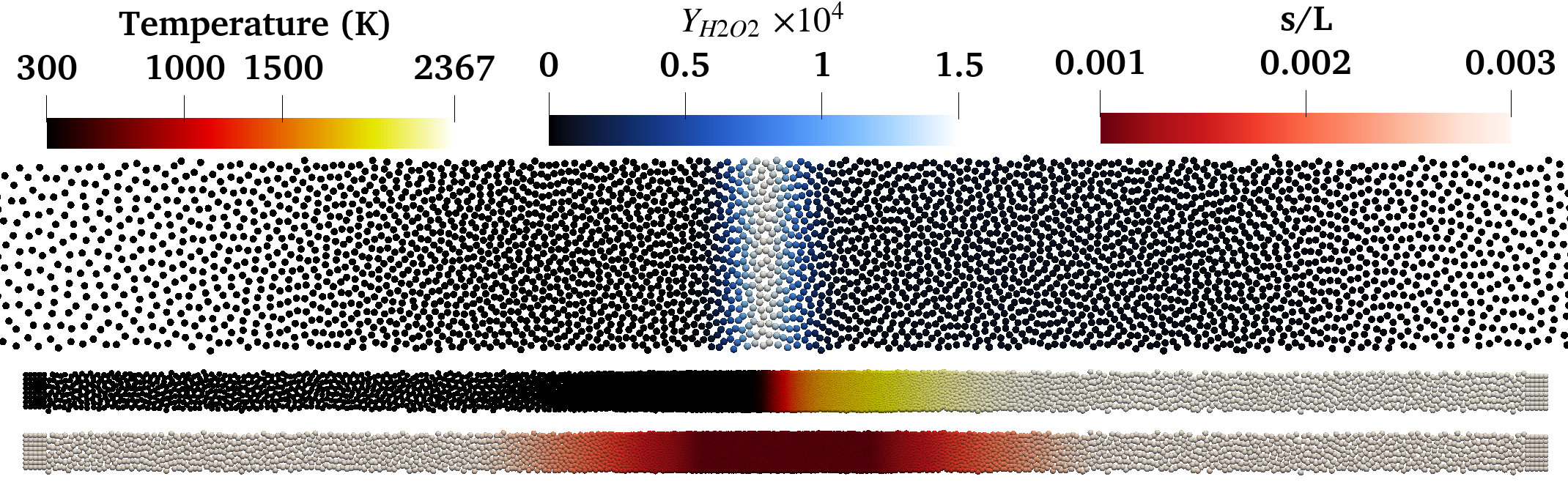}
\caption{The node distribution for the planar laminar hydrogen-air flame simulations. Dimensionless resolution $s/L$ (bottom), temperature profile (middle), and a magnified view of $H_{2}O_{2}$ mass fraction (top).\label{fig:flamedisc}}
\end{figure}

First I present the results of two-dimensional simulations of planar laminar freely-propagating stoichiometric hydrogen-air flames. The computational domain is rectangular, with length $L_{x_{1}}=L=10mm$ and a narrow aspect ratio with $L_{x_{2}}=L/40$. The domain is periodic in $x_{2}$, with inflow and outflow boundaries in $x_{1}$. In all cases, the temperature $T_{in}=300K$, velocity $U=2.05m/s$, and mass fractions for a stoichiometric hydrogen-air mixture are imposed at the inflow boundary, and a partially non-reflecting outflow condition tracks an outflow pressure of $p_{out}=10^{5}Pa$. I use the $9$ species, $21$ step reaction mechanism of~\cite{li_2004}. \hla{The initial temperature, density, velocity and mass fraction fields are defined by an error function flame profile with width $L/40$. The simulation is run until a steady flame profile is obtained.} Figure~\ref{fig:flamedisc} shows an example of the computational domain and node distribution for this test, with the bottom panel showing the resolution $s$, the middle panel showing the temperature, and the top panel showing a close up view of the node distribution around the flame front, coloured by the mass fraction of $Y_{H_{2}O_{2}}$. ${H_{2}O_{2}}$ is chosen for visualisation, as it has a very narrow distribution profile within the flame front. For this discretisation, I set $s/L=0.003$ at the inflow and outflow, reducing \hla{linearly over a region of $0.1L$} to $s/L=0.001$ \hla{in the central $0.1L$ of the domain}, which corresponds to a resolution of $s=10\mu{m}$. As can be seen in the top panel of Figure~\ref{fig:flamedisc}, this results in the structure of the flame being resolved by at least $10$ nodes.

\begin{figure}
\includegraphics[width=0.49\textwidth]{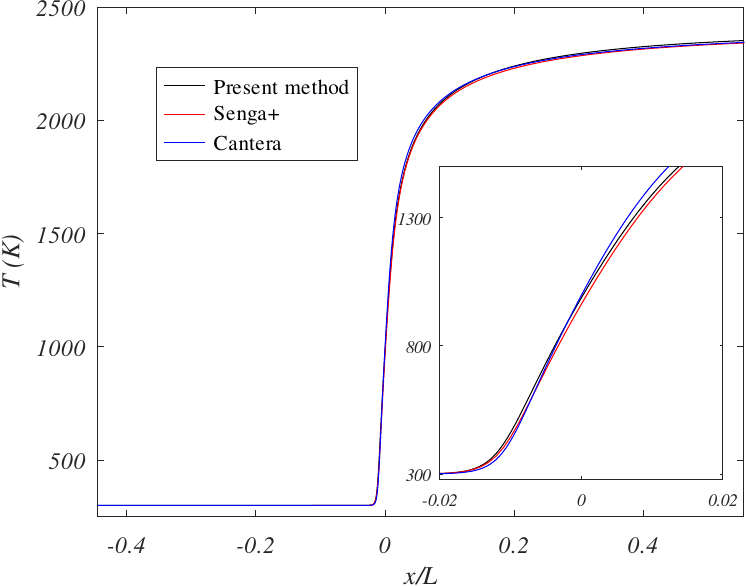}
\includegraphics[width=0.49\textwidth]{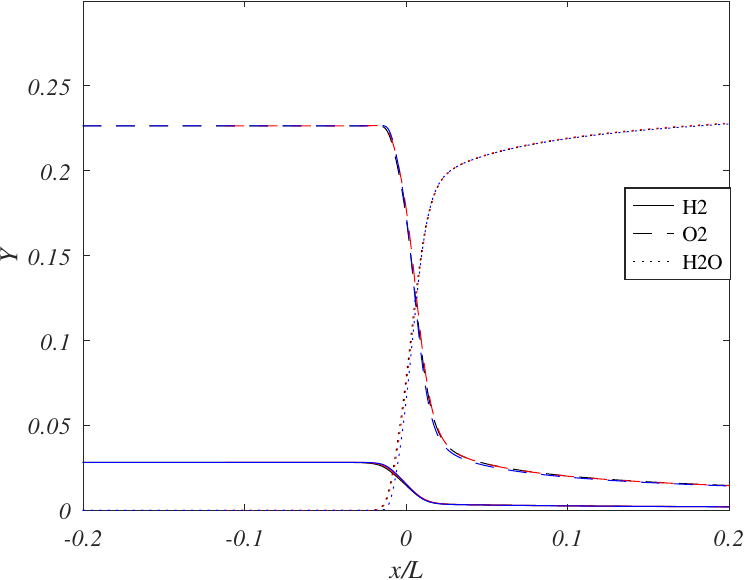}
\includegraphics[width=0.49\textwidth]{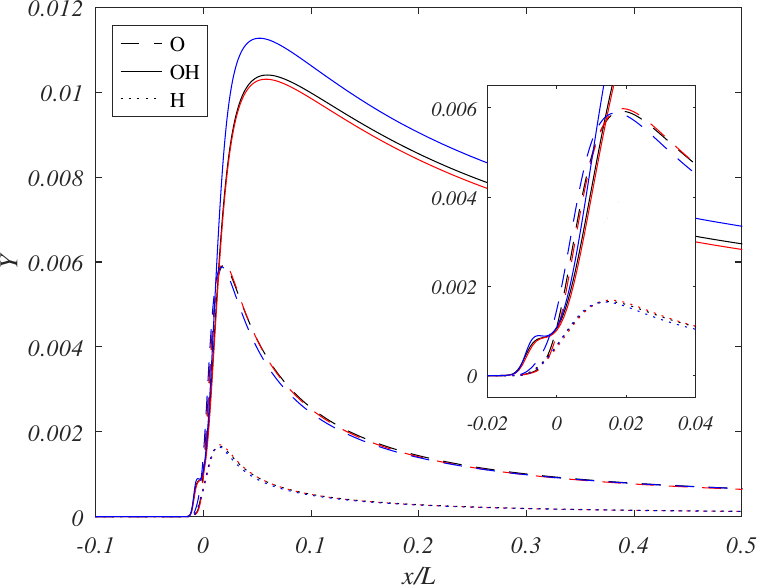}
\includegraphics[width=0.49\textwidth]{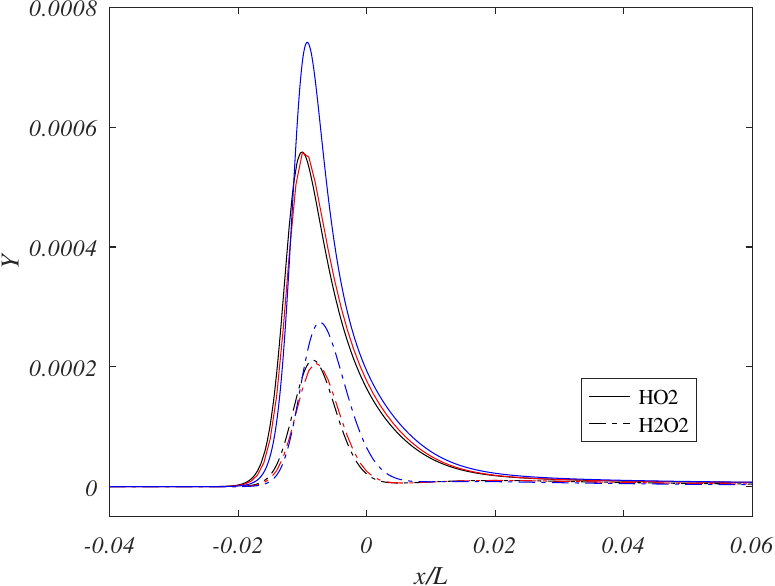}
\caption{The spatial profiles of temperature and species mass fractions through a freely propagating stoichiometric hydrogen-air flame, obtained with the present method (black lines), SENGA+~\cite{cant_2012} (red lines), and Cantera~\cite{cantera} (blue lines). For the present method and SENGA+, the 21 step mechanism of~\cite{li_2004} was used. For Cantera, the San Diego mechanism~\cite{sandiego} was used.\label{fig:h2_codecomp}}
\end{figure}

Figure~\ref{fig:h2_codecomp} shows the spatial profiles of temperature (top left) and species mass fractions through a freely propagating hydrogen-air flame, obtained with the present method (black lines), and two reference codes. \hlb{For the present method, the one-dimensional profiles are obtained by interpolating (using LABFM with $m=8$) values from the two-dimensional set of nodes to a uniform set of points along $x_{2}=0$.} The first reference code is SENGA+~\cite{cant_2012}, one of the combustion community's leading high-order DNS codes. SENGA+ is based on a tenth-order finite difference scheme, with a fourth-order explicit Runge-Kutta scheme for time integration. The set of equations solved by SENGA+ is very similar to those solved in my mesh-free framework, with only slight differences in the choice of combination rules for mixture-averaged transport properties. I also note there are differences between the present work and SENGA+ in the construction used for convective terms (SENGA+ uses a skew-symmetric form, the present work does not), and viscous terms. The simulations performed with SENGA+ use a uniform resolution of $10\mu{m}$. The second reference code is Cantera~\cite{cantera}, an open-source flame code, which in the present work is used to implicitly obtain steady one-dimensional flame solutions with an adaptive grid, and chemical reactions represented with the San Diego mechanism~\cite{sandiego}. To aid comparison, the data in Figure~\ref{fig:h2_codecomp} have been translated in $x_{1}$ such that the flame fronts (taken as the location where $Y_{H_{2}}$ is half its stoichiometric value) from all three codes align. 

In Figure~\ref{fig:h2_codecomp} there is a very close match between all three codes for the profiles of temperature and mass fractions of the main reactant and product species (top panels). For the $O$ and $H$ radicals, there is a very close match. However, for the mass fractions of $OH$, $HO_{2}$, and $H_{2}O_{2}$, the results from Cantera over-predict peak concentrations relative to both the present method and SENGA+. This may be due to the reaction mechanism used; with both my mesh-free method and SENGA+ I use the $9$ species mechanism of~\cite{li_2004}, whilst for Cantera, the San Diego mechanism~\cite{sandiego} is used. However, it could also be due to different formulations for molecular diffusion. The mixture-averaged formulation is used in all three codes, although there are differences in the combination rules used to evaluate $D_{\alpha}$, $\mu$ and $\lambda$ between codes. For example, in the present work the power-weighted average descibed in~\cite{ern_1994} is used for viscosity, whilst in SENGA+ the combination rule of~\cite{wilke_1950} is used. Additionally, in the present work, Soret and Dufour terms are neglected, whilst they are included in SENGA+. Whilst these differences in formulation appear to yield negligible differences in results for the stoichiometric hydrogen-air flames modelled here, for lean flames, they may be more significant. Notwithstanding these minor discrepancies with the solutions obtained from Cantera, with SENGA+ widely established as one of the leading codes for combustion DNS (see e.g.~\cite{domingo_2023}), the excellent match between the present method and SENGA+ in Figure~\ref{fig:h2_codecomp} provide a demonstration of the comparable ability of LABFM to simulate accurately simulate laminar flames.

\begin{figure}
\includegraphics[width=0.49\textwidth]{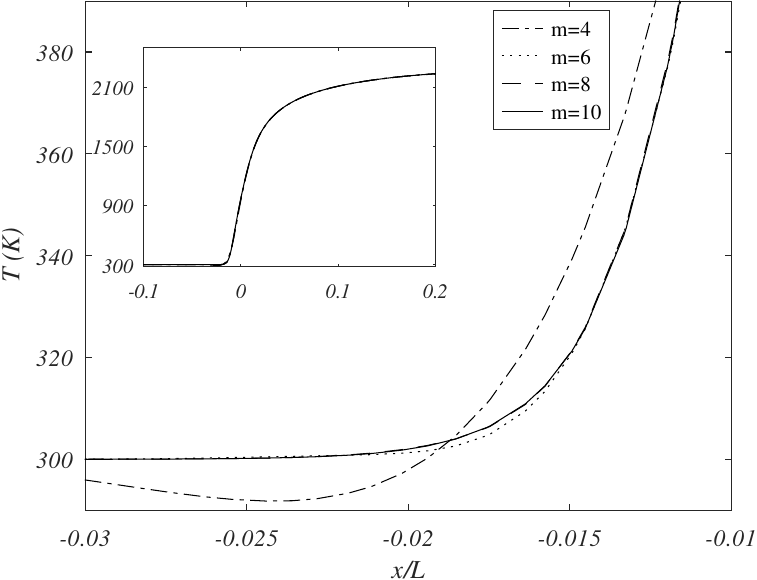}
\includegraphics[width=0.49\textwidth]{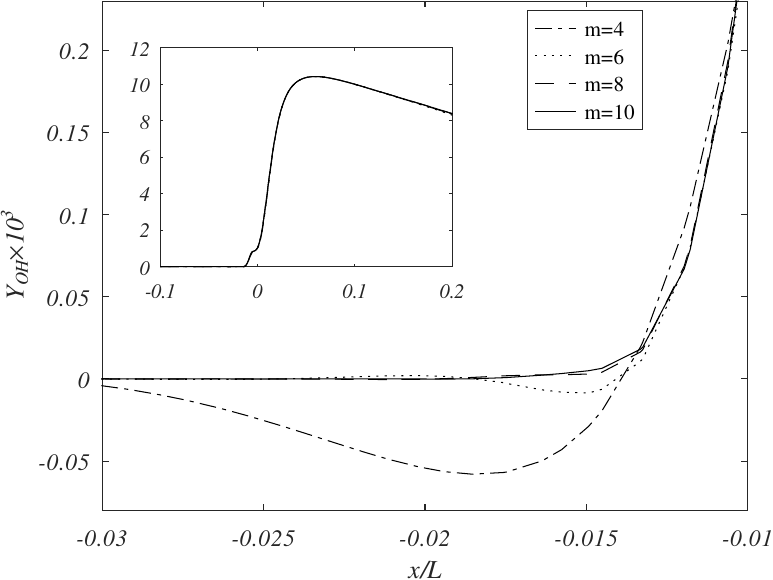}
\caption{The spatial profiles of temperature (left) and $OH$ mass fraction (right) through a freely propagating stoichiometric hydrogen-air flame, for different orders of LABFM: $m=4$ (dash-dot), $m=6$ (dotted), $m=8$ (dashed), and $m=10$ (solid). For all values of $m$, the simulation was run with a uniform resolution of $s=10\mu{m}$. Note $Y_{OH}$ has been scaled by a factor of $10^{3}$. \label{fig:h2_order}}
\end{figure}

I briefly explore the effect of the order of accuracy of my scheme, by simulating the same case, but changing $m$. Figure~\ref{fig:h2_order} shows the effects of changing $m$ on the profiles of $T$ and $Y_{OH}$. In all cases, the resolution is uniform at $s=10\mu{m}$. I test four orders of LABFM: $m=4,6,8,10$. For both the temperature (left panel) and $OH$ mass fraction (right panel) profiles, the results are indistinguishable when looking at the entire flame (inset in both panels). However, there are localised differences just upstream of the flame front (main axes). For smaller $m$ the resolving power of the scheme is lower, and the resolution is insufficient to accurately calculate the advection terms. This results in slight oscillatory behaviour just upstream of the flame front, with local minima in temperature, and a region of (unphysical) negative mass fraction of $OH$. For $m=4$, these are clearly visible in Figure~\ref{fig:h2_order}, whilst for $m=6$ the magnitude of these oscillations is much reduced. For $m\ge8$, these oscillations are not present - the temperature increases monotonically through the flame, and $Y_{OH}\ge0$ everywhere. I conducted simulations with an in-house one-dimensional finite difference code, constructed to solve the same set of equations as my mesh-free framework, but using high-order central finite differences instead of LABFM, and found the same behaviour with changing order of the scheme. \hlb{Note that whilst negative mass fractions are unphysical, that the numerical method admits them is not a fundamental problem, and is common to this and other methods (e.g. central finite differences) which are not strictly monotonicity preserving. In such schemes (including the present method), negative mass fractions are avoided by using high-order and sufficient resolution.}

\begin{figure}
\includegraphics[width=0.49\textwidth]{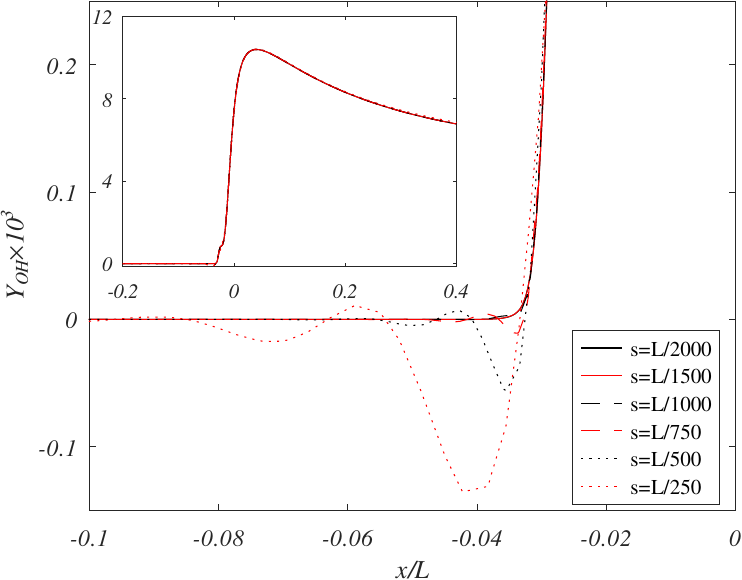}
\includegraphics[width=0.49\textwidth]{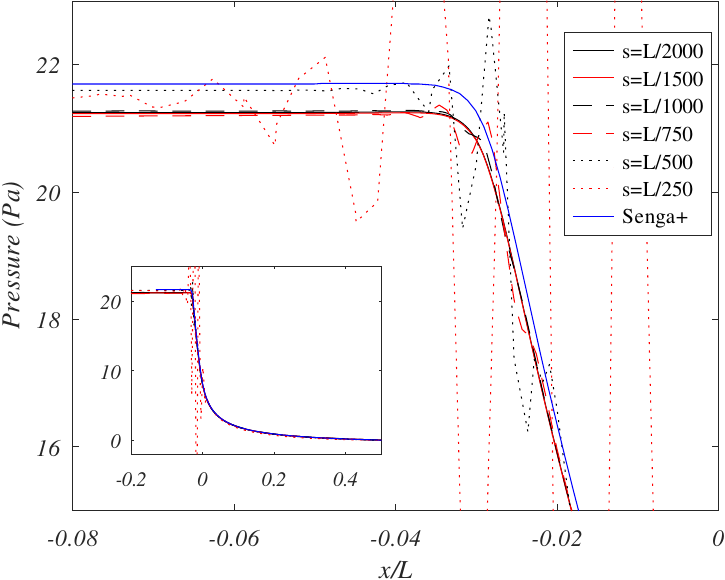}
\caption{\hla{Left panel:} the spatial profile of $OH$ mass fraction through a freely propagating stoichiometric hydrogen-air flame, for different resolutions. \hla{Right panel: the spatial profile of pressure (relative to the outlet pressure) across the flame, for the same set of resolutions, with comparison to the results from SENGA+ (blue line).} For all resolutions, the simulation was run with $m=8$. Note $Y_{OH}$ has been scaled by a factor of $10^{3}$.\label{fig:h2_res}}
\end{figure}

Next I keep $m=8$ and vary the resolution. In this test I set $s$ to be uniform throughout the domain. \hla{The left panel of }Figure~\ref{fig:h2_res} shows the profile of $Y_{OH}$ for a range of resolutions, from $s=L/250$ to $s=L/2000$, with the full flame shown in the inset, and a close up of the preheat zone in the main axes. \hla{The right panel of Figure~\mbox{\ref{fig:h2_res}} shows the profile of the pressure field through the flame, for the same range of resolutions, alongside data obtained with SENGA+ (blue line).} For $s\le{L}/1000$ the solution is converged, and discrepancies are between these finer resolutions are negligible. Conducting a simulation with a very fine resolution of $s=L/4000$ as a reference, I find discrepancies in the volume integrated heat release rate and the volume integrated kinetic energy are less thant $5.1\times10^{-4}$ for all $s\le{L}/750$. For larger $s$, small errors appear just upstream of the flame front, in the form of localised oscillations, similar to those observed in Figure~\ref{fig:h2_order} for $m=4$, with $Y_{OH}$ becoming locally negative. The magnitude of these errors increases with increasing $s$ (decreasing resolution). This observation is expected, and common to codes based on high-order central finite differences. It is a consequence of the discretisation scheme, resulting from the treatment of the convective terms: the scheme is not total variation diminishing - there is no upwinding or similar monotonicity preserving treatment. As in my investigation of changing $m$, I have performed additional simulations with an in-house finite difference code, which yielding the same oscillatory behaviour ahead of the flame for equivalent resolutions. \hla{As with the mass fraction profiles, the pressure field shows good agreement with SENGA+ for resolutions $s\le{L}/1000$, both in terms of the pressure jump across the flame (to within $0.5Pa$), and the profile through the flame. For coarser resolutions oscillatory behaviour is observed in the pressure field, and the magnitude of the jump across the flame changes too. These pressure oscillations are a consequence of the oscillations in the evolved mass variables ($\rho$, $\rho{u}_{i}$, $\rho{E}$ and $\rho{Y}_{\alpha}$): the the temperature is obtained via the non-linear system described in Section~\mbox{\ref{ge}}, and the pressure follows from the equation of state. I note that even for the coarse resolution of $s=L/500$, the maximum deviation in pressure is of the order of $3Pa$, which corresponds to a relative deviation (given atmospheric pressure of $10^{5}Pa$) of $3\times{10}^{-5}$.}

The results of Figures~\ref{fig:h2_order} and~\ref{fig:h2_res} taken together demonstrate the benefits of high-order schemes for flame simulations. With a high-order scheme, equivalent accuracy can be obtained at lower resolution (and hence cost). 

Finally, in addition to the results for hydrogen-air flames presented here, I have tested the method for stoichiometric methane-air flames, using the $16$ species, $25$ step reaction mechanism of~\cite{smooke_1991}. In the interests of brevity I omit detailed results for the methane-air flame, but note that I obtain qualitatively similar results: i.e. close agreement with Cantera for the major species, and very close agreement with results from SENGA+ for all species.

\subsection{Two-dimensional bluff-body stabilised flame}\label{sec:fh}

With the ability of the present method to accurately simulate planar laminar flames established, I now move on to a more challenging test. The key benefit of the present method over many other high-order combustion DNS codes is the ability to handle complex geometries, and as a demonstration of this, I simulate a two-dimensional bluff-body stabilised flame. Although comparatively simple (a single obstacle in a channel), this geometry is beyond the reach of most high-order combustion DNS codes. Similar configurations have been previously studied by a number of authors, both in the context of low-order codes (e.g. Ansys Fluent~\cite{fan_2014,vance_2019}), low-Mach number codes (e.g.~\cite{kedia_2014}), and high-order finite difference codes~\cite{lee_2015,kim_2018,kim_2019,kim_2021,kim_2021b}. I note that although the simulations in~\cite{lee_2015,kim_2018,kim_2019,kim_2021,kim_2021b} were performed with a high-order finite difference code, the geometric flexibility remains limited: in those works, the domain decomposition was constructed such that the bluff body boundary lay on the boundary between the domains of different processors. A bluff body with different shape (e.g. circular), could not be simulated with such an approach.

\begin{figure}
\includegraphics[width=0.49\textwidth]{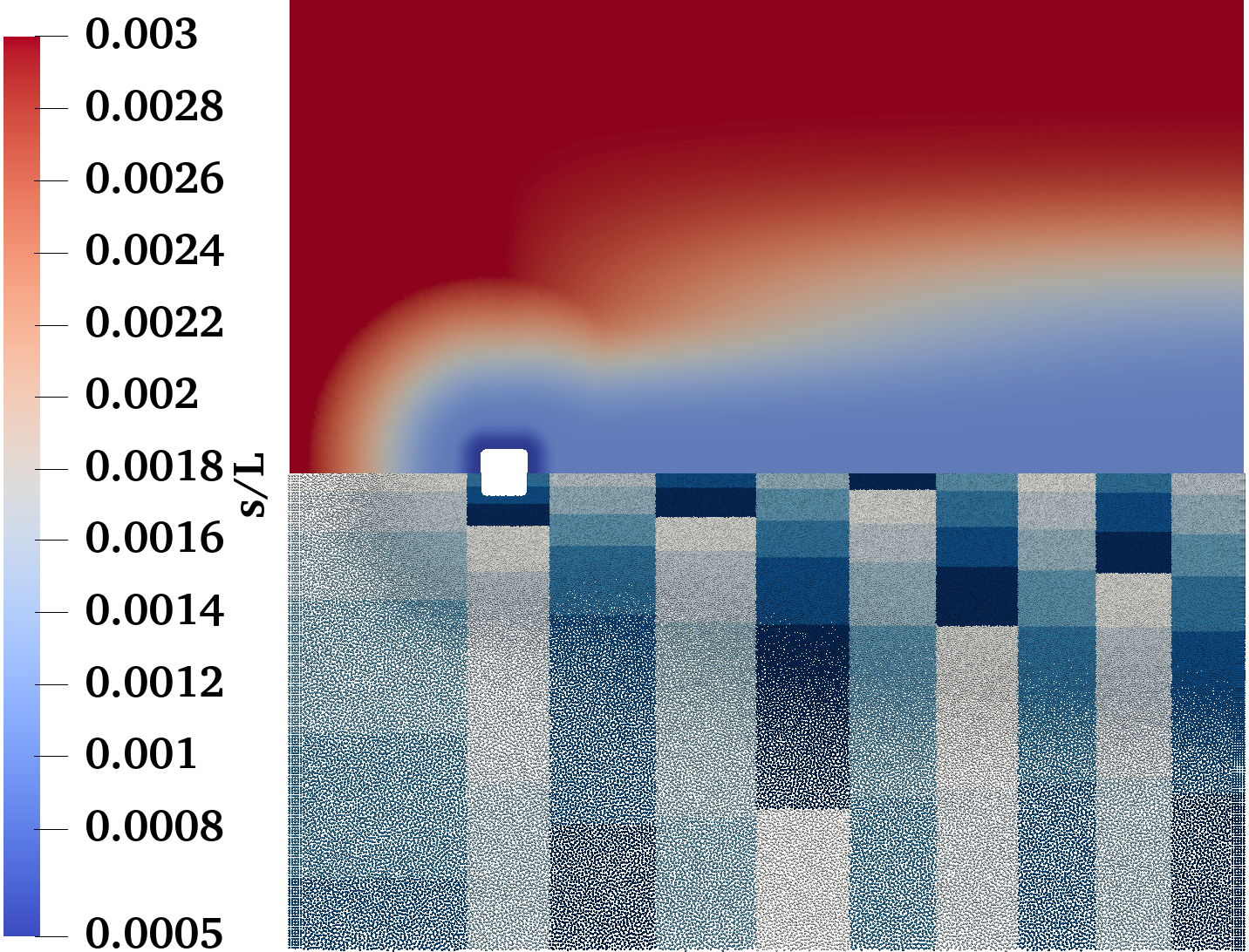}
\includegraphics[width=0.49\textwidth]{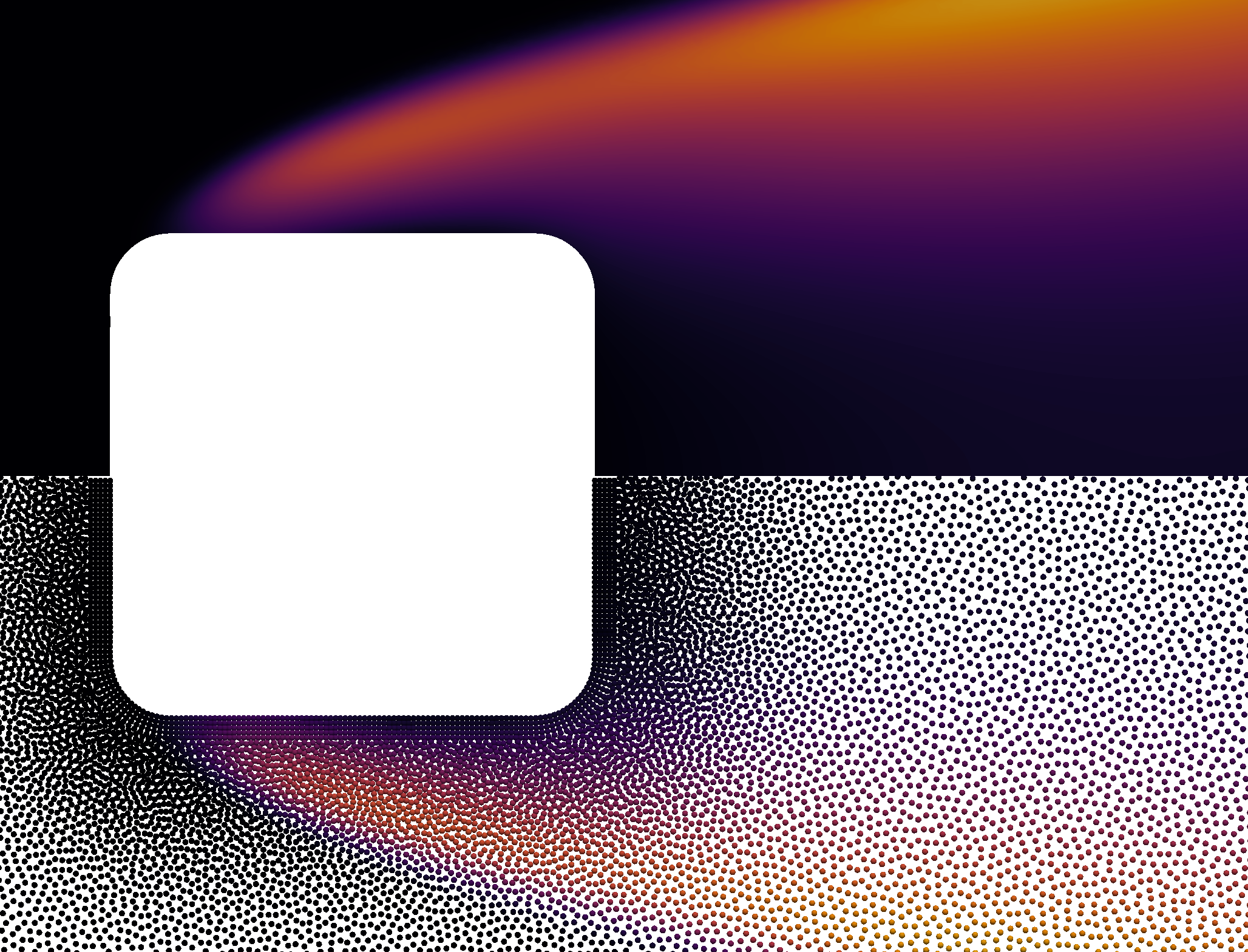}
\caption{Spatial discretisation for the two dimensional square bluff body configuration. Left, upper half: the resolution, normalised by the domain lengthscale $L$. Left, lower half: the domain decomposition with each coloured block representing a different MPI rank. Right: a close view of the bluff body, showing isocontours of the heat release rate (the colour level is a maximum at $10^{10}W/m^{3}$). The fields shown in the top half are interpolated onto a background mesh for visualisation purposes.\label{fig:square_rp}}
\end{figure}

For this test the configuration consists of a square two-dimensional domain with side length $L_{x_{1}}=L_{x_{2}}=L=10mm$. At the left and right boundaries non-reflecting inflow and outflow (respectively) conditions are imposed. The domain is bounded at the top and bottom by no-slip adiabatic walls. A bluff body is located with its leading edge a distance $L/5=2mm$ downstream of the inlet, and positioned centrally in the cross-channel direction. I consider two shapes of bluff body. First, a square body of side length $d=L/20=0.5mm$, on which the corners have been rounded to have a radius of curvature of $d/8$, and second, a circular body of diamter $d=0.5mm$. A no-slip adiabatic condition is imposed on the surface of the bluff body. The corners of the square bluff body are fully resolved with a resolution of $s_{min}=5\mu{m}$, whilst for the circular bluff body I set $s_{min}=10\mu{m}$. The domain is discretised with a non-uniform resolution with $s=s_{min}$ at the bluff body surface, increasing to $s=s_{max}=30\mu{m}$ away from the bluff body, in regions where combustion is not expected to occur. In regions combustion is anticipated to occur (i.e. near the bluff body, and directly downstream), the resolution is restricted to $s\le10\mu{m}$. This resolution matches that used in~\cite{kim_2019}, and ensures the flame front is resolved by at least $15$ nodes. \hla{Although no closed analytic expression is available to describe the variation of $s_{a}$ with $x_{1}$ and $x_{2}$, input files containing lists of nodes and resolutions are available on request.} The entire domain is discretised with approximately $2\times{10}^{5}$ nodes, a significant reduction compared to the $10^{6}$ nodes required for a uniform resolution of $s=10\mu{m}$, and $4\times{10}^{6}$ required for a uniform resolution of $s=5\mu{m}$. The simulations are run on $200$ cores spread across $100$ MPI ranks. The domain, node distribution, MPI decomposition, and resolution are illustrated in Figure~\ref{fig:square_rp}. 

Following~\cite{kim_2019}, the inflow composition is a hydrogen-air mixture with an equivalence ratio of $\Phi=0.5$, and the inflow temperature is set at $T_{in}=300K$. The partially non-reflecting outflow boundary is set to track a pressure of $p_{out}=10^{5}Pa$. The simulations is initialised with zero velocity, and ignition triggered by superposition of a Gaussian hot spot immediately downstream of the bluff body\hla{, with size $0.25mm$, and peak temperature $2500K$}. The inflow velocity is then set to a fully developed (i.e. parabolic) profile with mean velocity which ramps linearly from zero to $U=30ms^{-1}$ over a period of $T_{ramp}=L/2U$. For $U=30ms^{-1}$, the Reynolds number based on the bluff body diameter of $Re=810$. Simulations at sequentially higher $Re$ are performed, each starting from the previous simulation, and again increasing the inflow velocity linearly over a period of $T_{ramp}=L/2U$. 

\begin{figure}
\includegraphics[width=0.99\textwidth]{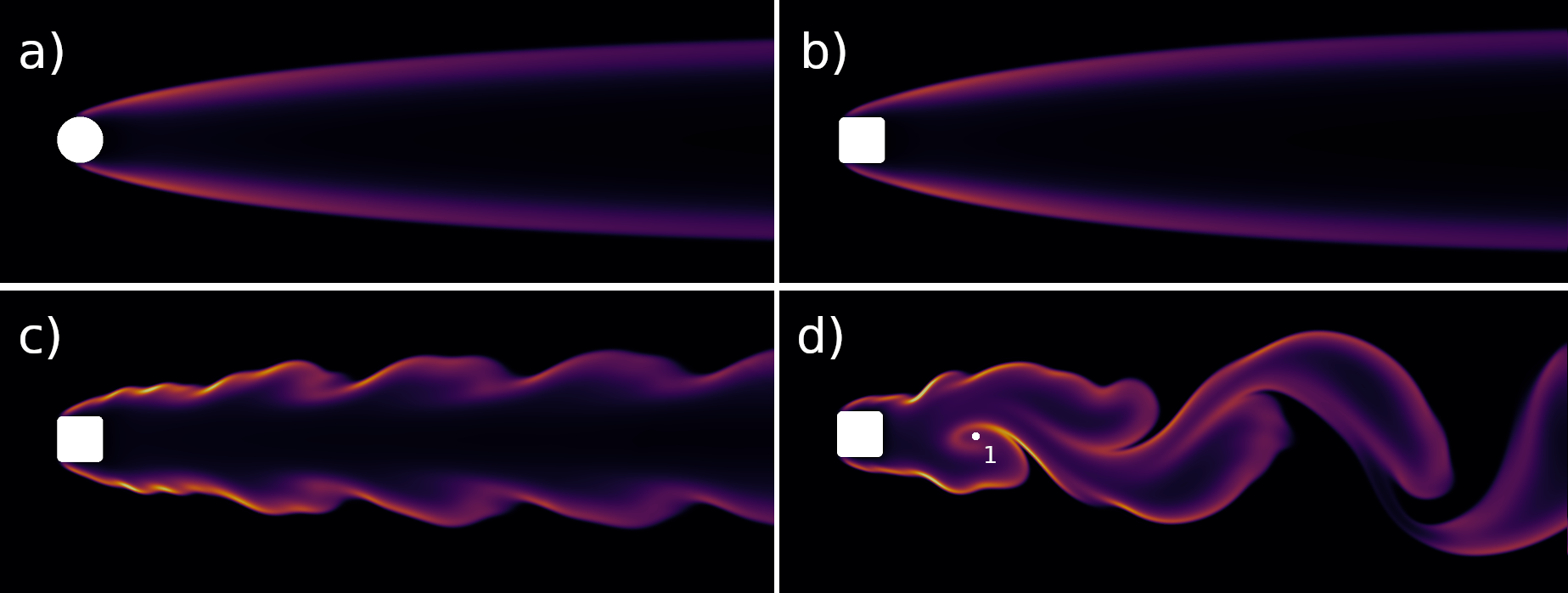}
\caption{Isocontours of the heat release rate for a hydrogen-air flame anchored to a bluff body. a) Circular bluff body with $Re=810$ and a steady flame; b)-d) a square bluff body with three flame regimes: b) steady, at $Re=810$, c) symmetric vortex shedding at $Re=2160$, and d) asymmetric vortex shedding at $Re=2160$. The colour level is a maximum at $2\times10^{10}W/m^{3}$).\label{fig:fh1}}
\end{figure}

\begin{figure}
\includegraphics[width=0.99\textwidth]{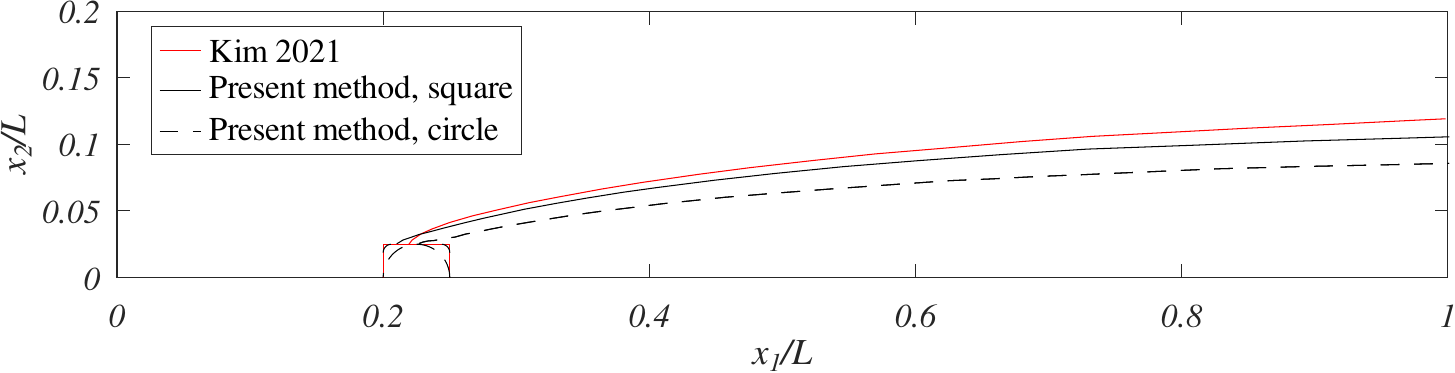}
\caption{Comparison of the steady flame shape with the present method (black) at $Re=810$, for the square (solid lines) and circle (dashed lines) bluff body, and the work of~\cite{kim_2021} (red lines, $Re=800$).\label{fig:fh2}}
\end{figure}

Figure~\ref{fig:fh1} shows the flame front obtained with the present method for several Reynolds numbers, illustrating the different dynamics. Panels a) and b) show a steady flame at $Re=810$ for the circular (a), and square (b) bluff body. Figure~\ref{fig:fh2} shows the geometry of these flames, as defined by the isocontour along which $Y_{H_{2}O}$ is half its equilibrium value, alongside the same measure taken from~\cite{kim_2021}. For the square bluff body (solid black lines) I see an approximate match with the data from~\cite{kim_2021} (red lines), but with a slightly reduced rate of expansion of the flame with increasing streamwise distance, and small differences around the flame attachement point. my method yields an attachment point slightly further upstream than in~\cite{kim_2021}, with a shallower (more swept back) profile near the bluff body. I postulate that the differences in attachement are due to the slight differences in geometry: my bluff body has rounded corners, whilst that in~\cite{kim_2021} is a square. The rounded corners on my geometry result in reduced flow separation at the sides of the bluff body, and in a higher energy boundary layer and hence the more swept back flame profile. The downstream differences in flame shape are likely to follow from the differences in the attachment. For the circular bluff body (dashed black lines), the flame attaches to the bluff body midway between the leading and trailing edges, and the expansion profile is shallower still. Although these differences are minor, they may have an impact on the dynamics of unsteady flames at higher $Re$, and serve to highlight the importance of simulations which accurately account for bluff body geometry. An in depth study of the effects of bluff body geometry on flame dynamics is planned for future work.

\begin{figure}
\includegraphics[width=0.99\textwidth]{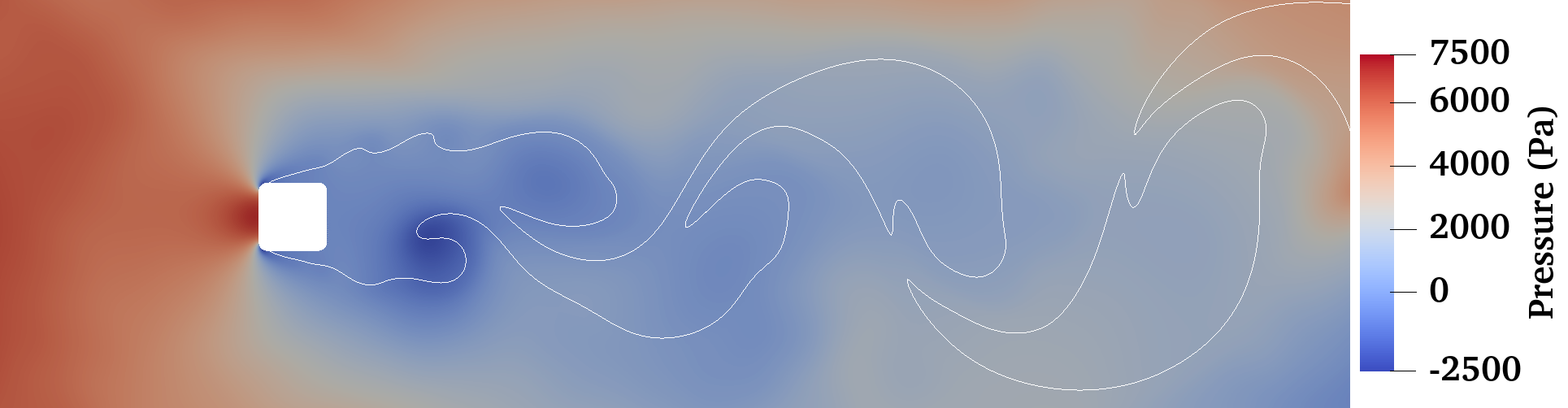}
\caption{Isocontours of pressure (relative to the outflow pressure) around a flame anchored to the square bluff body in the asymmetric vortex shedding regime. The contour at which $Y_{H_{2}O}$ is half its equilibrium value is shown in white.\label{fig:fh_press}}
\end{figure}

Panels c) and d) of Figure~\ref{fig:fh1} show the symmetric vortex shedding regime (c) and asymmetric vortex shedding (d), both at $Re=2160$ ($U=80ms^{-1}$), for the square bluff body. I see qualitatively similar dynamics to those presented in~\cite{kim_2018,kim_2019,kim_2021}. As the inflow velocity is increased, the steady flame develops mild fluctuations, which develop into symmetrically shed vortices, and then asymmetrically shed vortices. Regions of local extinction occur, and eventually these lead to global extinction. The mechanism appears to be the entrainment of cold gases into the wake of the bluff body due to the vortex shedding (near the point marked $1$ in panel d) of Figure~\ref{fig:fh1}), as observed in~\cite{kim_2019}. Whilst my simulations show flames with similar dynamics to those of~\cite{kim_2019}, I do not match their values of $U$ for the transition between regimes. In my work, I generally find the transition from steady to unsteady flames occurs at lower $U$ than that found in~\cite{kim_2019}. I believe this is due to the finite size of the domain. Whilst for the symmetric vortex shedding (and mild fluctuation) mode the fluctuations propagate convectively downstream, the instability appears to be an absolutely unstable one: that is, disturbances downstream can propagate upstream. The transition from a steady flame to the mildly fluctuating mode starts with small deviations from the steady flame far downstream from the bluff body, and the streamwise extent of these deviations then migrates upstream as they increase in amplitude. It is not clear whether this is due to the confinement (the flame is acoustically confined, even though hydrodynamically unconfined), or whether disturbances could propagate upstream even for a truly unconfined flame. Previous stability analysis of bluff-body stabilised flames (e.g.~\cite{anderson_1996,shanbhogue_2009}) has relied on assumptions of low Mach number, unconfined flames, and parallel flow, and open questions remain on the nature of this instability. Even for an acoustically unconfined flame, acoustic waves are partially reflected from the flame itself, and disturbances could potentially propagate upstream between the two flame brushes. \hla{Figure~\mbox{\ref{fig:fh_press}} shows the pressure field around the flame in the asymmetric vortex shedding regime. Whilst the high pressure at the leading edge stagnation point, and the low pressure in the recirculating vortex just behind the bluff body are visible, there is also significant large scale structure in the developed pressure field due to the interaction between the unsteady flame and the acoustic field.}

A further complication in this regard is the outflow boundary condition. Although able to handle the passage of an obliquely angled flame, the outflow boundary conditions (both in the present work, based on~\cite{sutherland_2003}, and in e.g.~\cite{kim_2019}, based on~\cite{yoo_2007}) provide imperfect (though very close) approximations to the incoming acoustic waves. The passage of a flame through the boundary generates a small amount of acoustic noise, and it is possible that this triggers the transition from a steady to an unsteady flame. This noise is expected to be greater for flames at an oblique angle than those perpendicular to the boundary, as the boundary condition is constructed to relax towards a state with zero tangential velocity. In the present case, the steady flame solution has non-zero $u_{x_{2}}$ within the flame, and without knowing this a priori, designing the boundary condition to relax towards $u_{x_{2}}=0$ on the boundary is, although not strictly correct, the best available option. Furthermore, the outflow boundary conditions cannot predict the flame dynamics in the far field. It is possible that in reality an instability can develop far downstream and propagate upstream, but this would not be captured in the present simulations with a finite domain.

\begin{figure}
\includegraphics[width=0.99\textwidth]{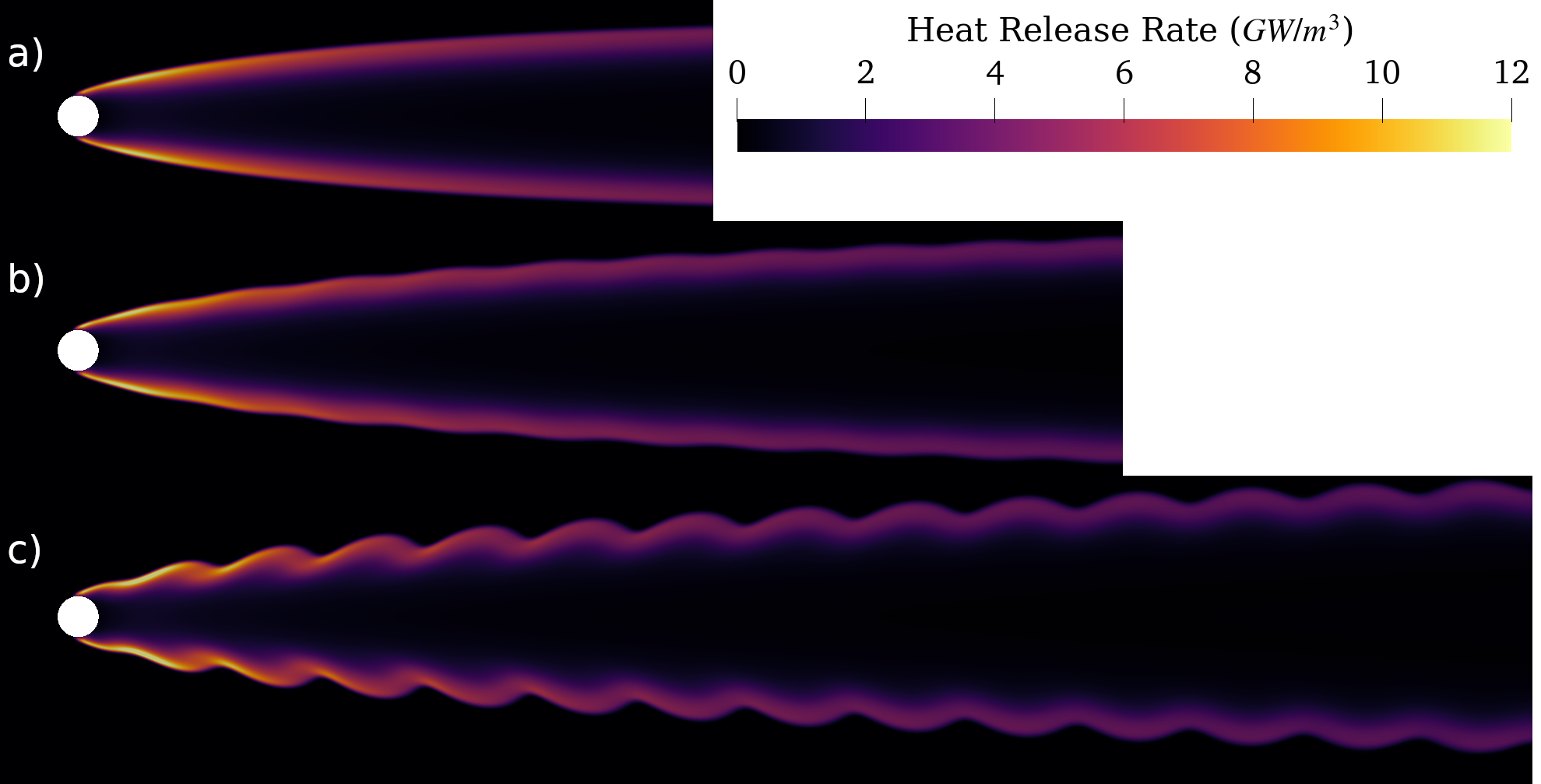}
\caption{Isocontours of the heat release rate for three domain lengths a) $L_{x_{1}}$, b) $1.5L_{x_{1}}$, and c) $2L_{x_{1}}$. In all cases, the simulations are started from the same initial conditions, and run until $t=3.75ms$.\label{fig:fh3}}
\end{figure}

I have conducted additional simulations with the circular bluff body, in which the length of the domain is increased in the $x_{1}$ direction by a factor of both $1.5$ and $2$. I find a dependence of the flame dynamics on the domain length. Figure~\ref{fig:fh3} shows isocontours of the heat release rate for flames attached to the circular bluff body at $Re=810$, with the upper a), middle b) and lower c) panels showing the domain lengths of $L_{x_{1}}$, $1.5L_{x_{1}}$, and $2L_{x_{1}}$ respectively. 
All three simulations are started with zero velocity, uniform composition, and a Gaussian hot spot just downstream of the bluff body to trigger ignition. In all cases the inflow velocity is increased linearly to reach $U=30$ at $t=0.25ms$. The results in Figure~\ref{fig:fh3} are taken for all three cases at $t=3.75ms$.

\begin{figure}
\includegraphics[width=0.49\textwidth]{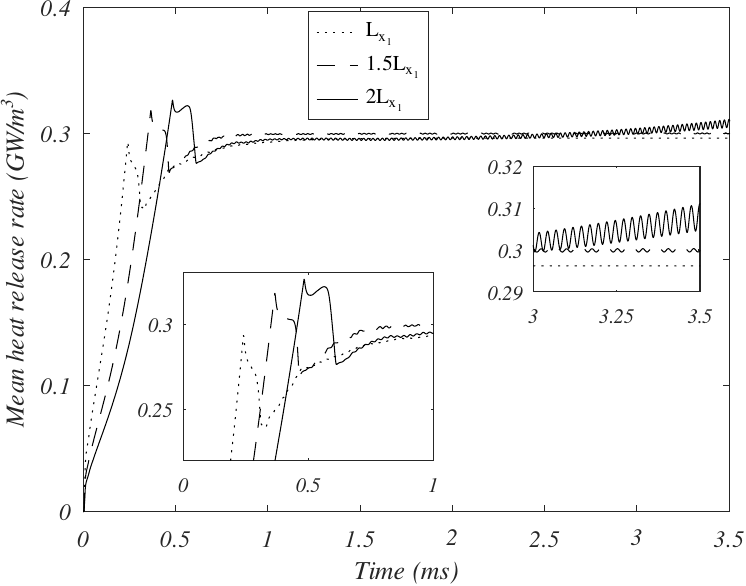}
\caption{Time evolution of the volume averaged heat release rate for simulations of the circular bluff body flame, for three domain lengths.\label{fig:fh4}}
\end{figure}

For the shorter domain the flame is stable: the flame profile shown in panel a) of Figure~\ref{fig:fh3} remains unchanged when the simulation is run for a further $5ms$. For the longest domain, the flame is unstable, developing mild fluctuations which move convectively downstream. For the domain with length $1.5L_{x_{1}}$, the flame is close to the stability limit; perturbations grow in time, but very slowly. This can be seen quantitatively in Figure~\ref{fig:fh4}, which shows the time evolution of the volume averaged heat release rate for the three domain lengths. In all cases, the heat release rate starts at zero, and ramps up as ignition occurs and the flame develops. The rate of increase is different for the three domain lengths, because of the normalisation by the domain volume. As the flame reaches the outflow boundary, some perturbations are triggered by the boundary conditions (visible in the lower left inset in Figure~\ref{fig:fh4}) for all three domain lengths. The behaviour of these perturbations at later times is shown in the upper right inset of Figure~\ref{fig:fh4}. For the short domain with length $L_{x_{1}}$, these perturbations are damped, and the volume averaged heat release rate reaches a steady value. For the longer domains these perturbations are less damped. For the domain with length $1.5L_{x_{1}}$, the perturbations grow, but extremely slowly, whilst the growth of the perturbations is clear for the domain with length $2L_{x_{1}}$. 

None of the above discussion implies that the unsteady dynamics shown in Figure~\ref{fig:fh1} (or in~\cite{kim_2018,kim_2019,kim_2021}) are unphysical or necessarily incorrect. Indeed, once the unsteady flame modes have been triggered, the boundary conditions are sufficient to capture the dynamics, which evolve convectively. It is only critical values of $U$ at which the dynamics transitions from steady to unsteady which are uncertain. It is not clear whether this is due to the construction of the boundary conditions, or the finite domain length, or both. Further work is needed in this area, and is ongoing within the author's research group.

\subsection{Inert isotropic decaying turbulence}\label{sec:tg}

I next demonstrate the ability of the method to handle inert turbulence via the benchmark problem of the three-dimensional Taylor-Green vortex. This problem has been widely used as a benchmark for high-order numerical methods~\cite{vanrees_2011,wang_2013,jacobs_2017,cant_2022}, and reference data associated with~\cite{wang_2013} is available online~\cite{hocfd1}. The computational domain is cubic, periodic in all dimensions, with side length $2\pi$. The domain is discretised with a uniform resolution, with unstructured nodes $\mathcal{A}_{12}$ (a section of the node distribution in $\mathcal{A}_{12}$ is shown in the left of Figure~\ref{fig:disc2}), and a regular distribution along the $x_{3}$ axis.

\begin{figure}
\includegraphics[width=0.49\textwidth]{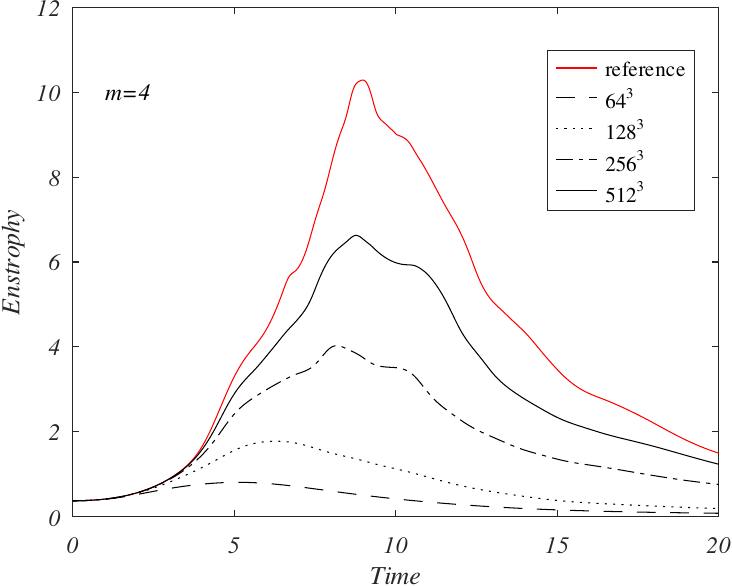}
\includegraphics[width=0.49\textwidth]{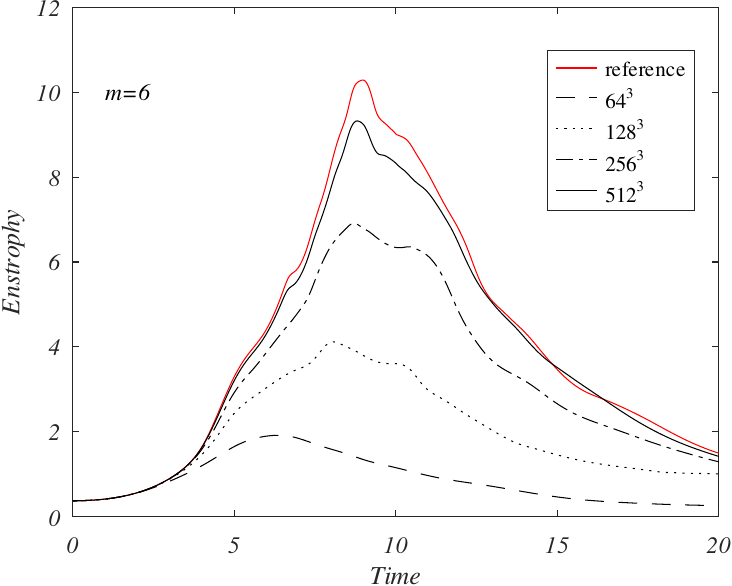}
\includegraphics[width=0.49\textwidth]{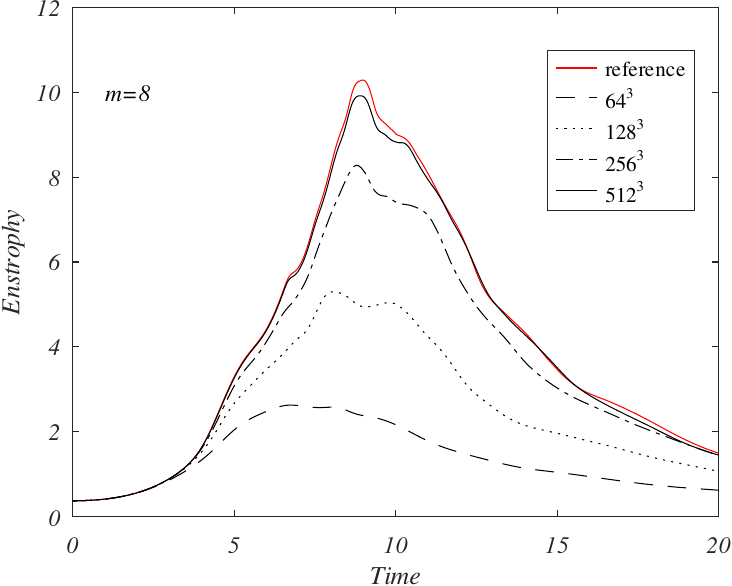}
\includegraphics[width=0.49\textwidth]{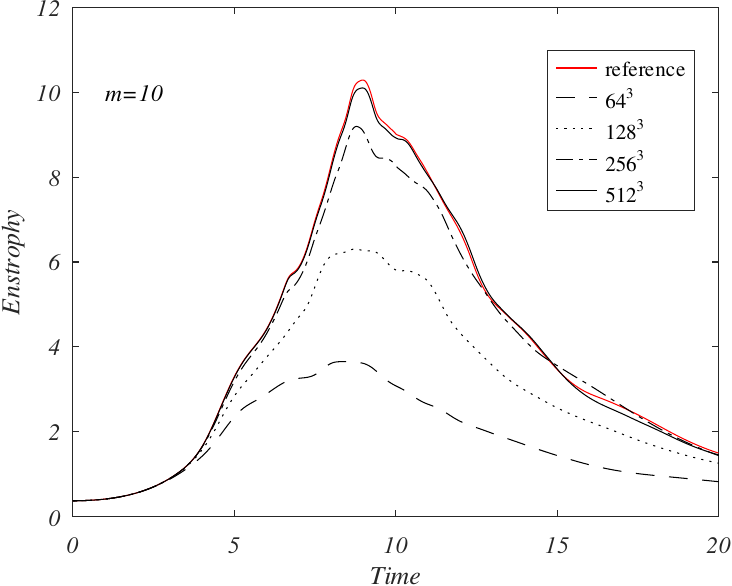}
\caption{The variation of volume averaged enstrophy with dimensionless time for the three-dimensional Taylor-Green vortices at $Re=1600$, for different orders of accuracy (top left - $4^{th}$ order, top right - $6^{th}$ order, bottom left - $8^{th}$ order, bottom right - $10^{th}$ order). In all plots the red line shows the reference data accompanying~\cite{wang_2013}.\label{fig:enstrophy}}
\end{figure}

\begin{figure}
\includegraphics[width=0.49\textwidth]{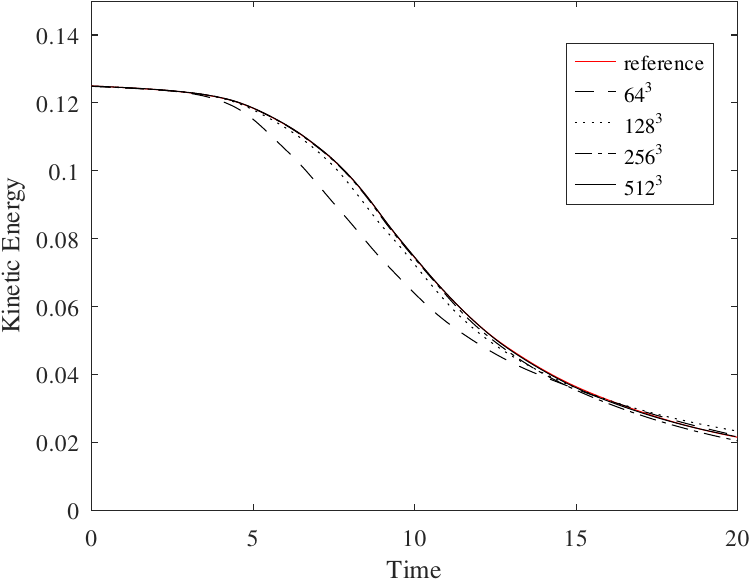}
\caption{The variation of volume averaged kinetic with dimensionless time for the three-dimensional Taylor-Green vortices at $Re=1600$, for $m=8$ and various resolutions. The red line shows the reference data accompanying~\cite{wang_2013}.\label{fig:ke}}
\end{figure}

The initial conditions are those given in~\cite{wang_2013} for compressible flows (sinusoidal and divergence free, but details omitted here for brevity), with an initial Reynolds number of $Re=1600$, and a Mach number of $Ma=0.1$. For this inert flow, I remove the temperature dependence of the transport properties by setting $r_{T}=0$, set $c_{p}$ to a constant independent of temperature, and $Pr=0.71$. Denoting the resolution by the number of nodes (i.e. $N=\left(2\pi/s\right)^{3}$), I systematically vary the resolution from $64^{3}$ to $512^{3}$, and the order of the scheme from $m=4$ to $m=10$. Figure~\ref{fig:enstrophy} shows the evolution of volume averaged enstrophy with time for this case, for this range of resolutions and orders. Figure~\ref{fig:ke} shows the time evolution of kinetic energy with $m=0.8$, for four resolutions (black lines). In both figures, the red lines correspond to the reference data accompanying~\cite{wang_2013}. 

My results approach the reference solution with both increasing resolution, and increasing polynomial consistency $m$. For $m=8$ and $m=10$, the maximum errors in enstrophy are small (below $3.6\%$ and $1.8\%$ respectively) at a resolution of $512^{3}$. For $m=4$, the error in the enstrophy deviates significantly from the reference solution even at the finest resolution. This is because the filter used for de-aliasing is only $4^{th}$ order, and the scales on which it acts are too close to the length scales of the flow. In terms of the kinetic energy evolution with $m=8$ (Figure~\ref{fig:ke}), the match is closer still, with only the coarsest resolution showing significant deviation from the reference data. The size of the computational stencil (and hence the cost) increases with $m$, and this has guided the choice in this work to generally opt for $m=8$ as a compromise between accuracy and efficiency. I note that a promising approach is to use a $m=4$ or $6$ for the spatial discretisation, but combined with $m=8$ for the filter. This will provide the cost benefit of reduced stencil sizes for the bulk of the calculation, whilst the higher order filter will not be overly dissipative at the physical scales of the solution. Furthermore, such an approach could easily be implemented adaptively, with $m=8$ in regions of interest (where fine flow structures or flames exist), and lower ($m=6$ or $m=4$) elsewhere.

These results demonstrate that the LABFM discretisation possesses sufficient accuracy for simulations of turbulent flows. \hlb{The discrepancies between the present method and the reference solution} are comparable to those of other leading methods for DNS, including the filtered finite difference schemes in~\cite{debonis_2013} and the high-order flux reconstruction scheme in~\cite{cant_2022}. \hlc{The discrepency compared with the reference solution is expected. The reference solution uses a pseudo-spectral method de-aliased by Fourier smoothing (see references within~\mbox{\cite{vanrees_2011}}), which exhibits negligible numerical dissipation up to about $80\%$ of the Nyquist wavenumber of the discretisation, whilst the LABFM operators have dissipation profiles similar to high-order finite differences, and for $m=8$, have negligible numerical dissipation only up to about $50\%$ of the Nyquist wavenumber, as shown in~\mbox{\cite{king_2022}}.} Furthermore, given the highly three-dimensional nature of this flow, they provide a validation of the coupling between the LABFM discretisation in $\mathcal{A}_{12}$, and high-order finite differences in $x_{3}$. I have performed additional simulations with the initial velocity field rotated by $90$ degrees, and obtain almost identical results in terms of the time evolution of enstrophy and kinetic energy, with the maximum discrepancies being $0.35\%$ and $0.057\%$ respectively.

\subsection{Flame-turbulence interactions\label{sec:fti}}

With the ability of the framework to simulate both flames and inert turbulence demonstrated, I briefly turn my attention to flame-turbulence interactions. I simulate the evolution of a statistically planar 3D flame under decaying isotropic turbulence. This problem has been studied extensively (see e.g.~\cite{trouve_1994,chakraborty_2005,han_2008, chakraborty_2009,dopazo_2017,keil_2021,awad_2022,cant_2022}) although often with slight differences in the detail, for example in the choices of turbulence intensity relative to laminar flame speed $u^{\prime}/S_{L}$, the turbulence lengthscale relative to flame thickness $l/\delta_{th}$, or the method by which the initial turbulence field is generated. 

The computational domain has size $L\times{L/2}\times{L/2}$, with $L=10mm$. The mean flow is in the $x_{1}$ direction, in which partially non-reflecting inflow and outflow conditions are applied, whilst the boundaries in the transverse directions are periodic. The resolution is uniform in the transverse directions, and varies from $s=L/125=80\mu{m}$ at the inflow and outflow, to $s=L/500=20\mu{m}$ at the centre of the domain where $x_{1}=L/2$. The resolution in $x_{3}$ is uniform, and set at the number-weighted mean value of $s$, giving $s_{3}=L/292=34.2\mu{m}$. In the interests of computational expedience, and to match existing simulations in the literature, the chemistry is represented by a single-step Arrhenius model tuned to match the laminar flame speed of a stoichiometric methane-air mixture, following~\cite{cant_2022}. Both reactions and products are assumed to have unity Lewis number, I set $Pr=0.7$, and take the temperature dependent exponent for the viscosity to be $r_{T}=0.7$. The flow is initialised from a one-dimensional laminar flame solution superimposed on a turbulent velocity field, with the unburnt gases having temperature $T_{u}=300K$, and the outflow tracking pressure $p_{out}=10^{5}Pa$. The turbulent velocity field is obtained by applying diffusion to a field of uncorrelated, uniformly distributed random numbers, following the method of~\cite{kempf_2005}. This approach is well suited to the present method where nodes are not arranged on a structured grid, and as the diffusion process is fully explicit, it is easily incorporated into the numerical framework. Furthermore, it is well suited to complex geometries, and the study of flame-turbulence interactions in non-trivial geometries is a key focus of future work. There are two key limitations of the method of~\cite{kempf_2005}. Firstly, the resulting velocity field is not divergence free~\cite{wu_2017}, which can result in non-negligible acoustic energy which must be dissipated in the early stages of the simulation. Secondly, it offers little control over the energy spectrum. By construction, the spectra of the velocity components are flat at small wavenumbers ($2\pi/l$), with sharp decay at higher wavenumbers. Despite these drawbacks, the method of~\cite{kempf_2005} is still used in turbulent combustion simulations, and I follow~\cite{cant_2022} in using it to generate the initial velocity field. The turbulent velocity field is homogeneous and isotropic, and I perform simulations at three different turbulent lengthscales $l/\delta_{th}\in\left[1,2,3\right]$ (where $\delta_{th}$ is the (laminar) thermal flame thickness). In all cases, I set the turbulent intensity $u^{\prime}=5S_{L}$, where $S_{L}$ is the laminar flame speed. With the flame located within the refined region, the thermal flame thickness $\delta_{th}\approx0.36mm$ is resolved by at least $10$ nodes. For this configuration, the Damk{\"o}hler numbers are $Da=1/5,2/5,3/5$ for $l/\delta_{th}=1,2,3$ respectively, and all three cases fall within the thin reaction zones regime. The simulation is run for two eddy turnover times $2l/u^{\prime}$. 

\begin{figure}
\includegraphics[width=0.9\textwidth]{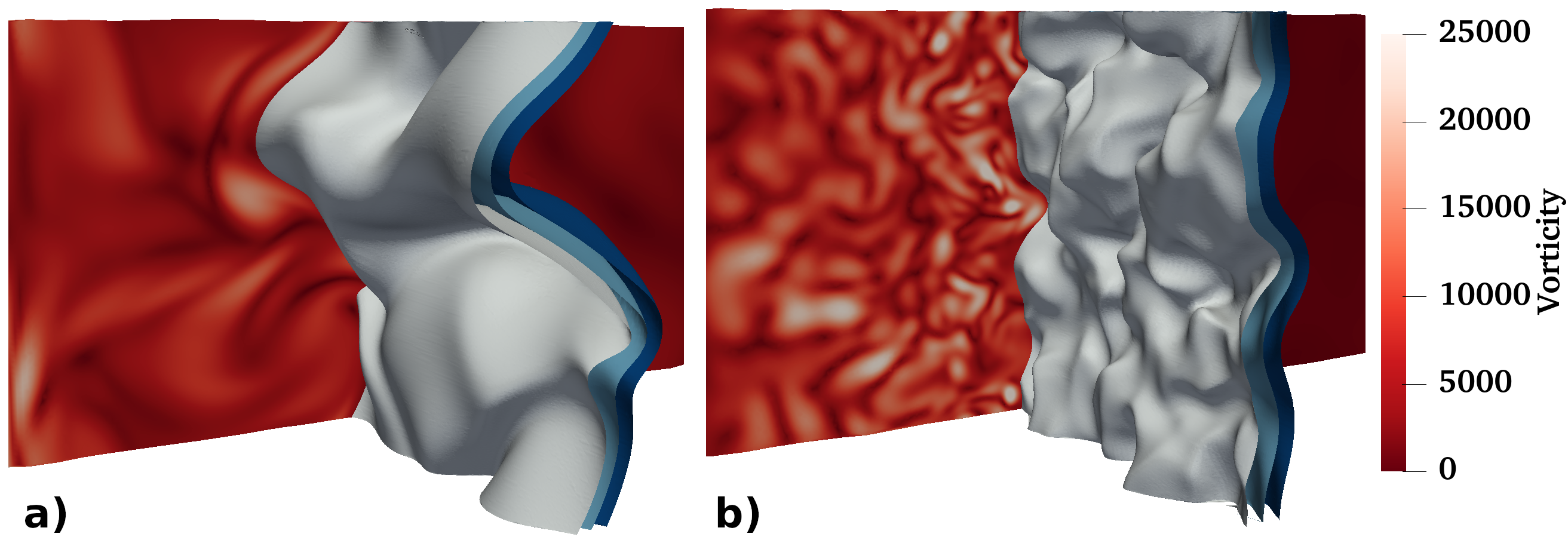}
\caption{Isosurfaces of the reactant mass fraction (in blue) at $Y_{R}=0.9$ (left), $Y_{R}=0.5$ (middle), and $Y_{R}=0.1$ (right) with slices showing contours of the vorticity magnitude (reds) for the flame-turbulence interaction test, for two values of turbulence length scale $l/\delta_{th}=3$ (a), and $l/\delta_{th}=1$ (b). In both cases the results correspond to $t=2l/u^{\prime}$.\label{fig:ft1}}
\end{figure}

Figure~\ref{fig:ft1} shows isosurfaces of the reactant mass fraction (in blue), over a background slice of the vorticity magnitude (in red), for two values of turbulence length scale: a) $l/\delta_{th}=3$ (left panel) and b) $l/\delta_{th}=1$ (right panel). In both cases, the results correspond to time $t=2l/u^{\prime}$. For $l/\delta_{th}=3$, the configuration matches that in~\cite{cant_2022}, and the resulting flame is qualitatively similar, whilst for $l/\delta_{th}=1$ the wrinkling of the flame is correspondingly smaller, both in typical lengthscale, and amplitude. For both cases, the vorticity magnitude downstream of the flame is suppressed at all scales due to the gas expansion across the flame, as predicted by theory~\cite{emerson_2012} and previous numerical experiments~\cite{keil_2021}.

\begin{figure}
\includegraphics[width=0.49\textwidth]{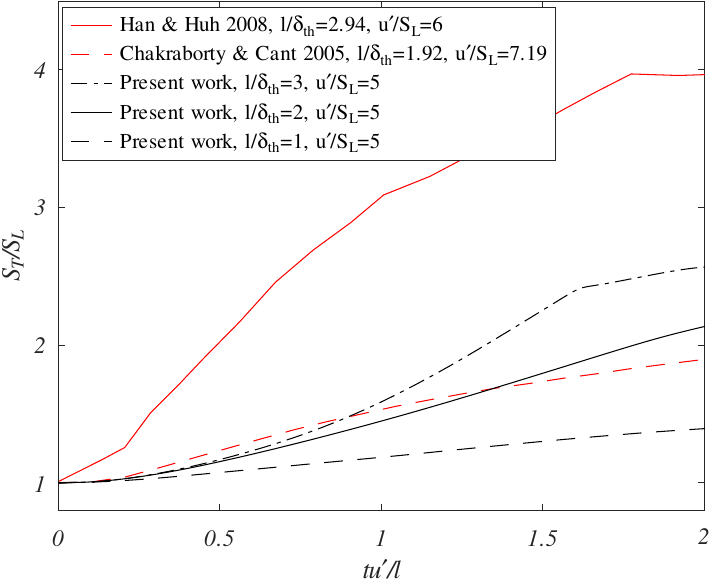}
\caption{Time evolution of the turbulent flame speed $S_{T}$ for two values of turbulent length scale $l/\delta_{th}=3$ (solid black line) and $l/\delta_{th}=1$ (dashed black line). The solid red line corresponds to data taken from~\cite{chakraborty_2005}, and the dashed red line corresponds to data taken from~\cite{han_2008}.\label{fig:ft2}}
\end{figure}

The turbulent flame speed is calculated by integrating the heat release rate over the domain, and the time evolution of the ratio of turbulent to laminar flame speeds $S_{T}/S_{L}$ is shown in Figure~\ref{fig:ft2}, alongside data from similar simulations taken from~\cite{chakraborty_2005} (solid red line) and~\cite{han_2008} (dashed red line). For $l/\delta_{th}=2$ (solid black line) I see a similar trend between the present simulations and the results of~\cite{chakraborty_2005}, with $S_{T}/S_{L}$ increasing by a factor of approximately $2$ over the first two eddy turnover times. For $l/\delta_{th}=1$ (dashed black line), the turbulent flame speed increases more slowly, and for $l/\delta_{th}=3$ (dot-dash black line) the turbulent flame speed increases more quickly, as expected. The discrepancy between the data of~\cite{chakraborty_2005} and~\cite{han_2008}, despite similar configurations, serves to highlight the challenge in accurate quantitative comparisons for this case. The evolution of the turbulent flame speed is strongly dependent on initial turbulent energy spectrum, as this has a significant bearing on how the turbulence develops over time. Whilst~\cite{chakraborty_2005} generated a divergence free initial velocity field in Fourier space, with a Batchelor-Townsend energy spectrum, my initial velocity field is not divergence free, and has a flat energy spectrum, although the lengthscale below which the energy rapidly drops off is similar. The results in~\cite{han_2008}, and my own numerical experiments, show some variability between individual DNS runs, so quantitative comparisons should be made on ensemble averages rather than the results of a single simulation. Nevertheless, the results shown in Figures~\ref{fig:ft1} and~\ref{fig:ft2}, especially when taken in the context of the results in Sections~\ref{sec:1d} and~\ref{sec:tg} indicate the present numerical framework is capable of accurately simulating flame-turbulence interactions.

\subsection{Hydrogen ignition in an array of cylinders}

Finally, I present the results of a simulation designed to show the potential of the numerical method for more complex geometries. I offer no experimental data or comparison with other numerical methods, and include this case simply as an example of the geometries for which the present method is suited. \hlb{Whilst the fundamentals of the numerical framework have been extensively analysed in previous papers~\mbox{\cite{king_2020,king_2022}}, and the ability of the method to simulate flames, turbulence, and flame turbulence interactions has been assessed above, this test introduces a step change in the level of complexity, and further assessment and validation will be conducted in an in-depth study in future.}

The computational domain has size $13.53D\times2.5D\times{D}$, with $D=1mm$. Non-reflecting inflow and outflow boundary conditions are imposed in $x_{1}$, with periodic conditions in $x_{2}$ and $x_{3}$. An array of cylinders each with diameter $D$ is located with the centre of the central cylinder a distance $2.25D$ downstream of the inlet. The cylinders are arranged in an isometric configuration, with spacing $1.25D$. The domain contains a stoichiometric hydrogen-air mixture, with inflow velocity $U=10ms^{-1}$, and inflow temperature $T_{in}=300K$. For the results reported herein, I use dimensionless time $t^{\star}=tU/D$. The outflow boundary tracks a pressure of $p_{out}=10^{5}Pa$. The domain is discretised with a non-uniform resolution, varying from $s_{min}=\left(5/6\right)\times{10}^{-5}m$ at the cylinder surfaces, to $s_{out}=2s_{min}$ at the outflow, and $s_{in}=6s_{min}$ at the inflow. \hla{No closed form analytical expression is available to describe the spatial variation of $s$ through the domain, but input files containing lists of node positions and resolutions are available on request.} This \hla{discretisation} ensures that in regions where the flame is expected to reside, the resolution is approximately the minimum sufficient for laminar flame simulations, as identified in Section~\ref{sec:1d}. i conduct both two- and three-dimensional simulations. In each case, an inert simulation is run for one flow-through time, to obtain an initial velocity field. The simulation is then restarted with a Gaussian hot spot (with peak temperature $T_{hot}=2500K$ and characteristic width $0.2mm$ for two-dimensional simulations, and $1mm$ for three-dimensional simulations), located just upstream of the cylinder array, a distance $2mm$ upstream of the central cylinder, and centered in the $x_{2}$ and $x_{3}$ directions. The hot spot triggers ignition of the mixture, and I simulate the flame propagating through the cylinder array and downstream. 

\begin{figure}
\includegraphics[width=0.49\textwidth]{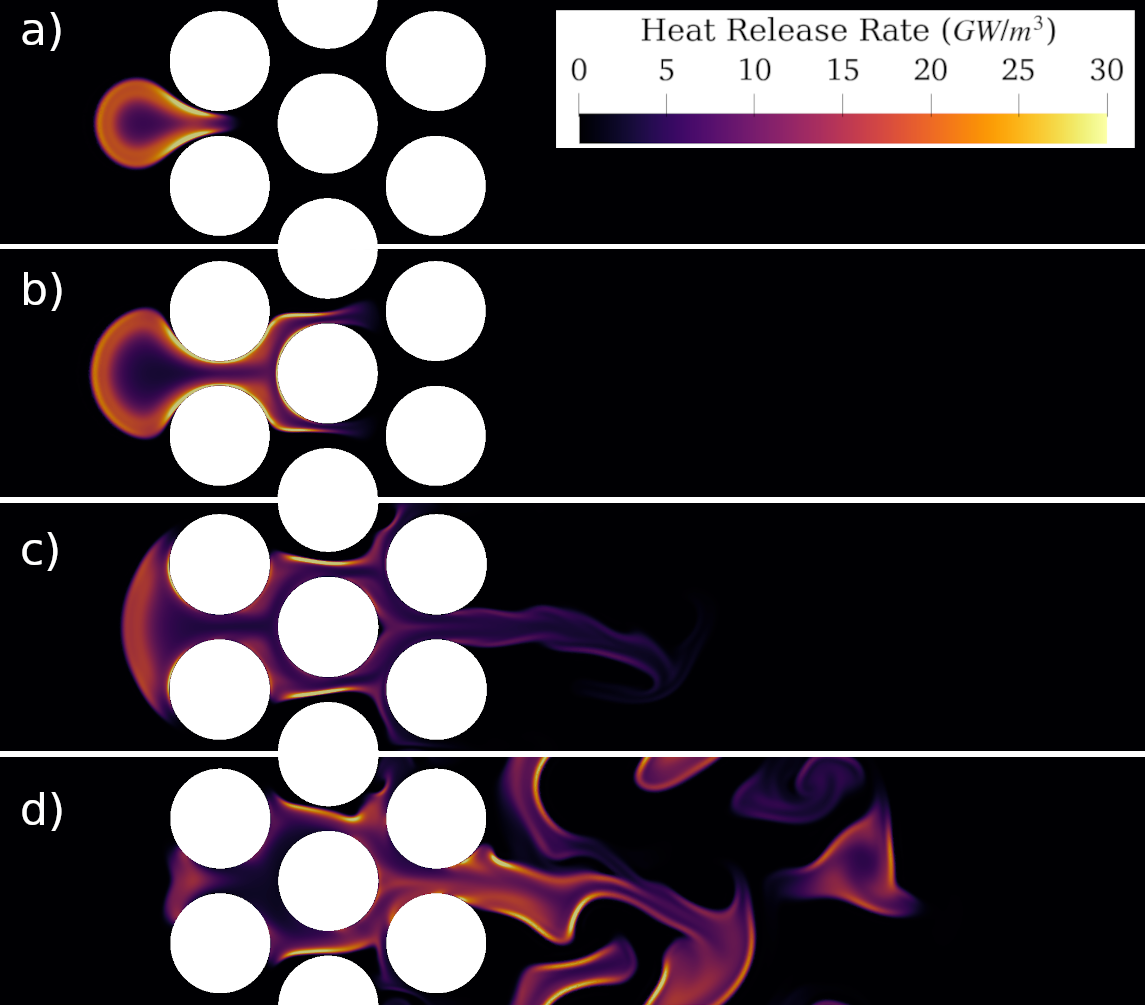}
\includegraphics[width=0.49\textwidth]{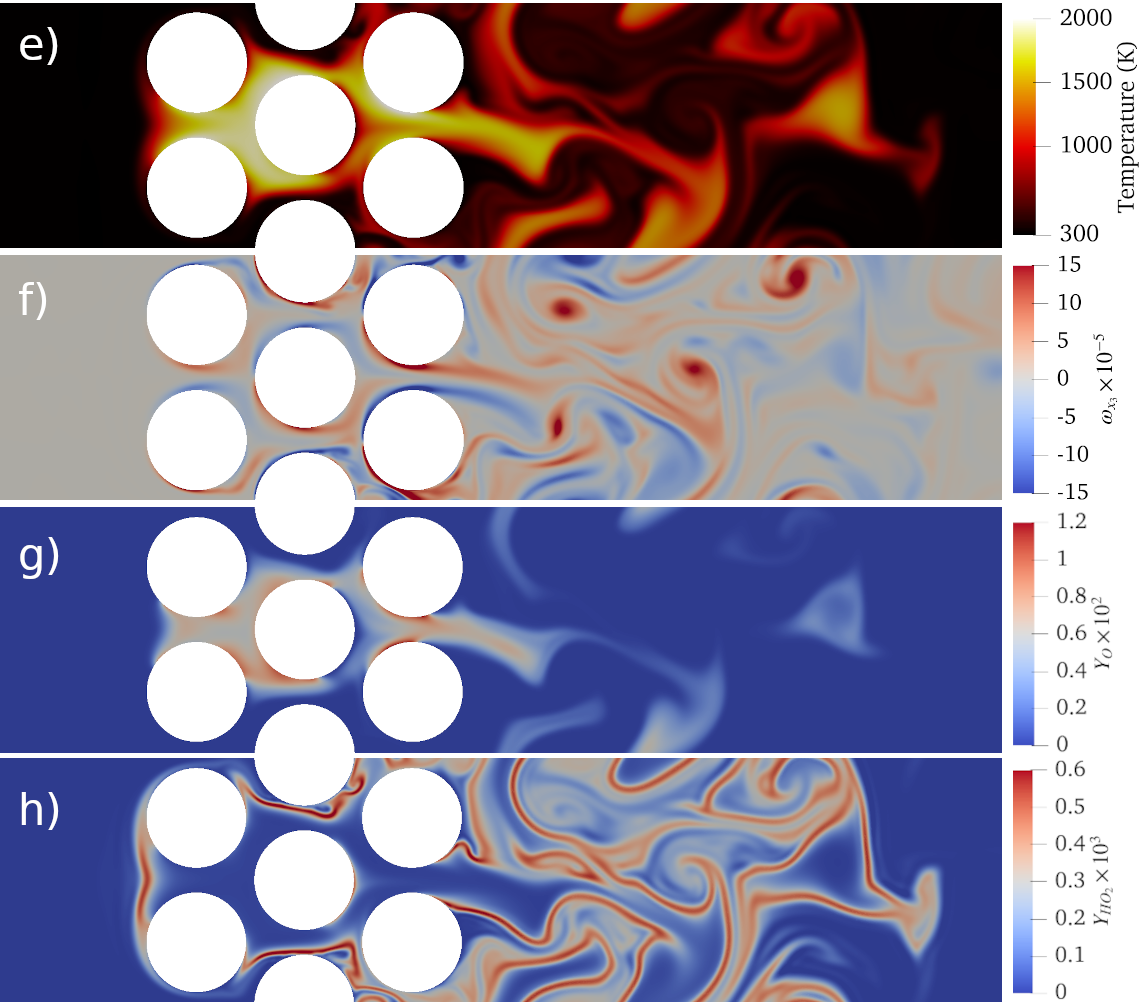}
\caption{Flame development for ignition in an array of cylinders, obtained from two-dimensional simulations. Panels a) to d) show isocontours of the heat release rate at dimensionless times a) $t^{\star}=0.2$, b) $t^{\star}=0.4$, c) $t^{\star}=0.8$, and d) $t^{\star}=1.25$ after ignition. Panels on the right show e) temperature, f) $x_{3}$ component of vorticity $\omega_{x_{3}}$ (scaled by $10^{-5}$), g) $Y_{O}$ (scaled by $10^{2}$), and h) $Y_{HO_{2}}$ (scaled by $10^{3}$), all at $t^{\star}=1.25$.\label{fig:porous1}}
\end{figure}

Figure~\ref{fig:porous1} shows the development of the flame kernel through the cylinder array for the two-dimensional simulation. The left panels a) to d) show isocontours of the heat release rate. The flame develops from the Gaussian hot spot and is initially drawn in to the array as can be seen in panels a) and b). A later times, the flame kernel begins to propagate through the cylinder array wake (panels c) and d) of Figure~\ref{fig:porous1}), at which point the flame speed increases due to the vortical structures in the flow. The right panels e) to h) show isocontours of different fields at dimensionless time $t^{\star}=1.25$. In panel f) the increase in secondary and unsteady vortical structures in the downstream half of the cylinder array is apparent. In panels g) and h), which show the mass fractions $Y_{O}$ and $Y_{HO_{2}}$ respectively, the assymetry of the flame near the leading edge of the cylinder array is visible. This can also be seen in the isocontours of the heat release rate in panel d). At this stage the flame doesn't connect with the upper and lower cylinders in the central column. I also see in panel g) a very non-uniform distribution of $Y_{O}$ within the flame in the cylinder array, with significant accumulations at certain locations on the cylinder walls, likely due to the small scale vortical structures within the array.

\begin{figure}
\includegraphics[width=0.99\textwidth]{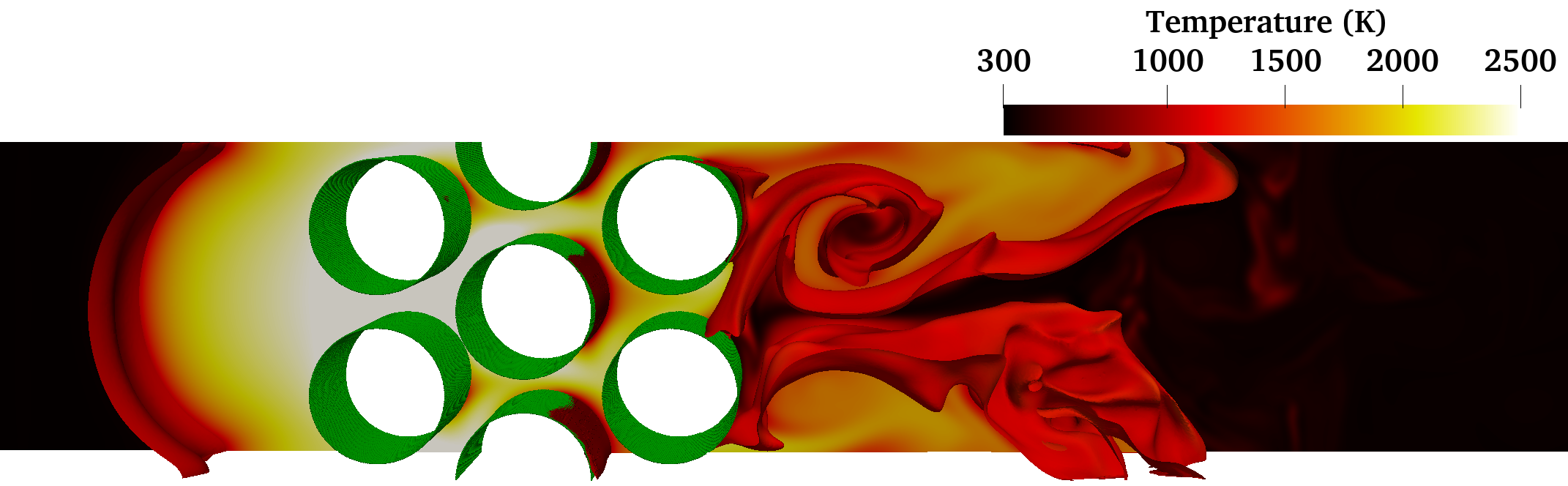}
\caption{The flame kernel at $t^{\star}=0.6$, obtained from a three-dimensional simulation. The flame is shown as an isocontour of the progress variable $\left(Y_{H_{2},in}-Y_{H_{2}}\right)/Y_{H_{2},in}=0.4$, where $Y_{H_{2},in}$ is the mass fraction of $H_{2}$ at the inlet. The cylinders are shown in green, whilst the flame and a background slice at $x_{3}=0$ show isocontours of the temperature.\label{fig:porous2}}
\end{figure}

Figure~\ref{fig:porous2} shows the flame kernel at time $t^{\star}=0.6$ for the three-dimensional simulation. Note, the hot spot used for ignition is larger for the three-dimensional simulation, and the flame develops faster, whilst the upstream flame-front has not yet been advected towards the cylinder array. The cold-flow velocity field is largely two-dimensional, with fluctuations in $x_{3}$ of $u^{\prime}_{x_{3}}\approx5\times10^{-3}U$. As the flame front develops, it triggers significant three-dimensional flow, especially in the vortical structures downstream of the cylinder array, and the flame kernel itself becomes highly three-dimensional in this region. At later times (not shown), the velocity fluctuations in the third dimension reduce and the flow becomes nearly two-dimensional again. 

Although I offer only observations of this test (I leave detailed analysis and further study of such problems to a future paper), it highlights the type of problem to which the present method is well suited: DNS of combustion in complex geometries, discretised to a high-order, and able to capture flame-vortex-structure (or flame-turbulence-structure) interactions. With such problems beyond the the capabilities of leading high-order finite difference codes for combustion DNS, the present method offers a complementary approach which extends the state-of-the-art for high-order methods in non-trivial geometries.

\section{Computational performance and scaling}\label{sec:cp}

\hla{As discussed in Section~\mbox{\ref{sec:intro}}, DNS of turbulent combustion is computationally demanding. The requirement to resolve the Kolmogorov scales gives rise to the number of collocation points scaling with $Re^{9/4}$~\mbox{\cite{domingo_2023}}, and the requirement to resolve the flame structure is often even more restrictive. With the present method having similar resolving power to high-order finite differences, the resolution requirements are similar to other leading combustion codes (e.g. SENGA+~\mbox{\cite{cant_2012}}, S3D~\mbox{\cite{chen_2011}} and KARFS~\mbox{\cite{perez_2018}}. Simulations at larger $Re$, although possible, require signicant increases in computational resources, and with the present work focussed on demonstrating the potential of the method, I have limited my investigations to flows up to $Re\approx{2000}$.}

\hlc{For the quasi-one-dimensional laminar flame tests conducted in Section~\mbox{\ref{sec:1d}}, a comparison with the two reference codes is not possible as for this test SENGA+ is used to perform truly one-dimensional simulations, whilst the present method is performing two-dimensional simulations, and Cantera (also truly one-dimensional) is solving for the steady state, rather than the unsteady system. However, comparisons can be made against other high-order finite difference codes. Broadly, the construction of the numerical framework is similar to other high-order finite difference codes for combustion (e.g. SENGA+), and the primary difference in computational costs arise from the difference in stencil sizes. For a finite difference scheme of order $m$, the computational stencil contains $\mathcal{N}=dm+1$ elements, where $d$ is the number of dimensions. For the present framework, the computational stencils are larger: for $m=6$, $\mathcal{N}\approx55$, and for $m=8$, $\mathcal{N}\approx75$ in two dimensions (and on average, as there is spatial variation in this due to resolution gradients). For three-dimensional simulations the values of $\mathcal{N}$ for the present method increase by $+m$ to include the finite difference stencil in $x_{3}$. Hence, for a two-dimensional simulation of order $m=8$, the computational stencil is approximately $4.5$ times larger for the present method than an equivalent finite difference scheme. This is the cost of geometric flexibility. The benefit, however, is that variable resolution can be easily implemented, and in a more flexible and localised manner than is possible with grid-stretching techniques for finite difference schemes.} 

\hlc{I take as an example the two-dimensional bluff body simulations in Section~\mbox{\ref{sec:fh}}. The simulations of~\mbox{\cite{kim_2019}} utilised a uniform grid with approximately $10^{6}$ collocation points. In the present work, the same geometry is discretised with approximately $2\times{10}^{5}$ collocation points, with a coarser resolution in regions where the flame is unlikely to occur (i.e. upstream of the bluff body, and away from the channel centre). For this case, the benefits of variable resolution (smaller $N$) and the costs of a mesh-free scheme (larger $\mathcal{N}$) approximately balance (to an order of magnitude), with~\mbox{\cite{kim_2019}} reporting costs of approximately $800$ CPU-hours per millisecond simulated, compared with $1132$ CPU-hours per millisecond for the present work. A caveat to this comparison is that the two sets of simulations use codes of differing maturity (a new code here, vs. one subject to extensive development and optimisation in~\mbox{\cite{kim_2019}}), run on different HPC systems, and with different approaches for certain aspects (e.g. use of Chemkin libraries, vs. in-house libraries in the present work). Furthermore, this is just one case, and the balance will be different in different scenarios: there will be cases where a uniform grid-based scheme is always faster, and cases where the present method performs better. There will also be cases which simply cannot be simulated by existing high-order finite-difference combustion codes, such as the final test case in this work, and these are the problems on which I intend to focus this method in future.}

\hlc{The memory requirements of the present method are greater than finite difference schemes (and more akin to high-order flux-reconstruction (FR) schemes), as a set of weights (or interpolant coefficients for FR schemes) must be stored for each collocation point (element/cell in FR schemes). This requirement is only prohibitive when performing scaling tests (e.g. three-dimensional Taylor-Green vortices at a resolution of $N=512^{3}$ running on $8$ processors). Although the number of elements in a computational stencil is larger than for a finite difference scheme, the stencil lengthscale $h/s$ is comparable. Hence, the amount of information which must be passed between subdomains is comparable with an equivalent order finite difference scheme (i.e. similar sized halos). As with high-order finite difference schemes, provided the number of nodes per MPI rank is large enough (typically $>6000$), the MPI communication does not present a bottleneck to simulation, with the cost of communication remaining small compared to the cost of calculation.}

\begin{figure}
\includegraphics[width=0.49\textwidth]{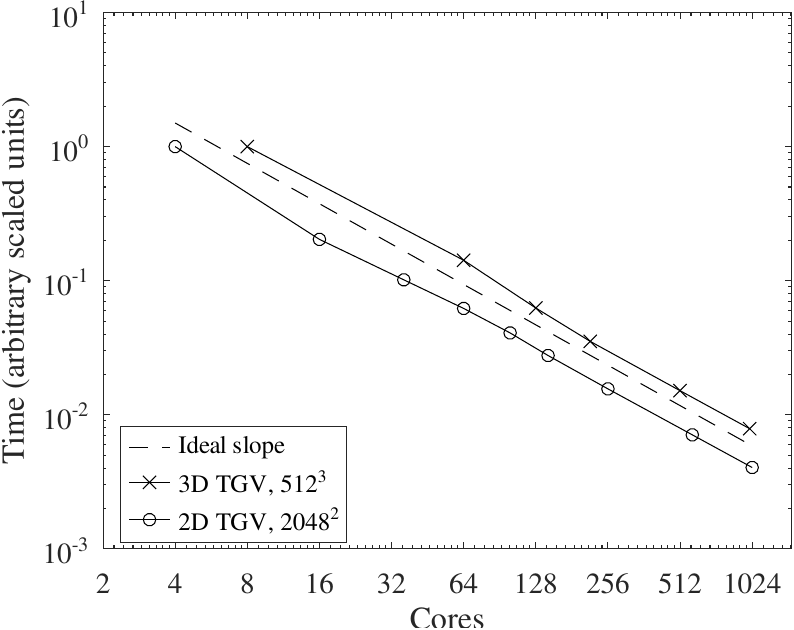}
\caption{Strong scaling of the code for the benchmark problem of decaying Taylor-Green vortices, in two- (circles) and three- (crosses) dimensions. The dashed line shows the slope for perfect scaling. Note that times have been re-scaled to arbitrary units.\label{fig:scaling}}
\end{figure}

\hlc{Figure~\mbox{\ref{fig:scaling}} shows the strong scaling performance of the method for the Taylor-Green vortex test case from Section~\mbox{\ref{sec:tg}}, for both two-dimensional (circles) simulations at a resolution of $2048^{2}$, and three-dimensional simulations at the finest resolution of $512^{3}$. In both cases, simulations were performed with a single thread per MPI rank, and so this is a test of the scaling performance of the distributed memory parallelism. In both cases there is almost perect scaling from the smallest number of cores ($4$ and $8$ in two and three dimensions respectively) to the largest available on the system ($1024$ and $1000$, for two and three dimensions respectively). For the two-dimensional case, the parallel efficiency from $4$ to $1024$ cores is $96.49\%$. Deviations above ideal scaling for the three-dimensional case at low numbers of cores ($<128$) are due to slight differences in configuration (e.g. not utilising every core on a compute node, to give each MPI rank enough memory.). For the three-dimensional simulations, the scaling efficiency is ideal. For the three-dimensional simulations run on $1000$ cores, there are approximately $1.342\times{10}^{5}$ collocation points per MPI rank, and the cost of MPI communication is small relative to the cost of calculations.}

\section{Conclusions}\label{conc}

Whilst direct numerical simulations (DNS) of combustion have long been restricted to simple geometries, the development of methods which provide the accuracy, speed and flexibility to enable the computation of flows in non-trivial geometries is an active area of research for the community. In this work I have presented a framework which allows for high-order discretisations of complex geometries. The discretisation is \emph{mesh-free} in two-dimensions, utilising the Local Anistropic Basis Function method (LABFM) to evaluate spatial derivatives, combined with high-order finite differences for the third dimension. 
The method is validated against several two- and three-dimensional test cases, yielding excellent agreement with a leading high-order finite difference code for laminar flame simulations. The ability of the method to handle turbulent flows is demonstrated with convergence towards reference solutions for simulations of a three-dimensional Taylor-Green vortex, whilst the invariance of the numerical results to rotations of the discretisation frame show the compatibility of the coupling of a mesh-free method in two dimensions to high-order finite differences in the third. My results show that increasing the polynomial consistency (and hence the resolving power) allows equivalent levels of accuracy to be obtained at coarser resolutions, and I find that using the method with consistency of order $m=8$ provides sufficient accuracy for resolving both flames and turbulent structures at reasonable resolutions. For flame-turbulence interactions, results are comparable with those in the literature, whilst close agreement with published simulations is obtained for bluff body stabilised flames, demonstrating the potential of the method for DNS of combustion in non-trivial geometries.

\hla{This new mesh-free method has sufficient resolving power for combustion DNS, and offers an alternative route to high-order combustion DNS in non-trivial geometries.} Whilst the present method can handle spatially varying resolutions, the discretisation is constant. A key advance which is under development is a framework for resolution adaptivity, such that the discretisation is dynamically refined in regions where the length-scales of flow features are small. Additionally, in depth studies are planned for bluff body stablised flames, and the effects of bluff body geometry on flame dynamics.

In the present work, the geometric complexity of the domain is limited to two dimensions, with homogeneity a requirement for the third due to the use of finite differences in that dimension. I note that there is nothing precluding the use of LABFM for all three dimensions, although the size of the computational stencil will be significant, increasing costs. Developments are under way create optimised stencils, which still yield high-order interpolants, but at reduced costs. This will render a three-dimensional implementation of LABFM more feasible, allowing truly complex geometries to be simulated.

\section*{Acknowledgements}
This work was funded by the Royal Society via a University Research Fellowship (URF\textbackslash R1\textbackslash 221290), and by a Dame Kathleen Ollerenshaw Fellowship awarded by the University of Manchester. I would like to acknowledge the assistance given by Research IT and the use of the Computational Shared Facility at the University of Manchester. I am grateful to Dr Steven Lind for many enjoyable discussions on mesh-free methods. I also express my thanks to Prof. Stewart Cant and Dr Umair Ahmed for providing (and supporting the use of) the code SENGA+.

\appendix

\section{Interpolation scheme for boundary discretisation}\label{ap:ifd}

Here I describe in detail the interpolated finite difference scheme used for nodes $a_{0}$ and $a_{1}$ near wall boundaries. For this appendix, I use vector notation for simplicity of exposition, with $\bm{r}=\left(x_{1},x_{2},x_{3}\right)$ denoting a position vector, \hlb{and the reader is referred to the diagram in Figure~\mbox{\ref{fig:interpfd}} for the definitions of the nodes $a_{0}$ to $a_{4}$}. Whilst the original form of LABFM in~\cite{king_2020} was constructed to obtain approximations to derivatives of a function, it can be easily modified to interpolate function values at specific locations, simply by including an additional zeroth-order element in the construction of the consistency matrices, and defining the interpolation operator as
\begin{equation}L_{a}\left(\phi\right)=\displaystyle\sum_{{b}\in\mathcal{N}_{a}}\phi_{b}w_{ba},\label{eq:interpo}\end{equation}
where the weights $w_{ba}$ are designed such that $L_{a}\left(\phi\right)$ approximates the value of $\phi$ at the position $\bm{r}_{a}$. Note that within the summation only $\phi_{b}$ appears (rather than $\phi_{ba}$ as in the original formulation), and hence~\eqref{eq:interpo} has no explicit dependance on $\phi_{a}$, and so may be used to evaluate $\phi$ at locations where no node exists. In the context of interpolation within the boundary stencil used herein (and illustrated in panel a) of Figure~\ref{fig:interpfd}), the interpolation operator \hlb{is used to evaluate properties at the fictitious nodes $a_{3}$ and $a_{4}$, and} takes the form
\begin{equation}L_{\chi}^{\zeta}\left(\phi\right)=\displaystyle\sum_{{b}\in\mathcal{N}_{\chi}}\phi_{b}w^{\zeta}_{b\chi},\label{eq:bdo}\end{equation}
for $\chi=a_{0},a_{1}$ and $\zeta=a_{3}, a_{4}$. Here $\chi$ indicates the (real) node over the stencil of which the interpolation is performed, and $\zeta$ indicates the fictitious node at which $\phi_{\zeta}$ is evaluated.
Here the weights $w^{\zeta}_{b\chi}$ are, as in the usual LABFM, obtained by solving a local consistency matrix, although in this case zeroth order monomials, and a zeroth order basis function (a constant) are included, resulting in the rank of the consistency matrix increasing by one.

The stencil for interpolation is centred on node $a_{2}$ \hlb{as illustrated in Figure~\mbox{\ref{fig:interpfd}}}, and the weights are constructed as
\begin{equation}w^{\zeta}_{b\chi}=\bm{W}\left(\bm{r}_{ba_{2}}\right)\cdot\bm{\Psi}_{\chi}^{\zeta}\end{equation}
where $\bm{W}\left(\bm{r}_{ba_{2}}\right)$ is the vector of basis functions. \hlb{To define the basis functions I first define a companion vector of doublets $\left(q_{1},q_{2}\right)$, and then for the series of doublets}
\begin{equation}\bm{q}=\left[\left(0,0\right),\left(1,0\right),\left(0,1\right),\left(2,0\right),\left(1,1\right),\left(0,2\right),\left(3,0\right),\left(2,1\right)\dots\right]^{T}\end{equation}
\hlb{the corresponding element of $\bm{W}\left(\bm{r}_{ba_{2}}\right)$ is given by}
\begin{equation}W\left(\bm{r}_{ba_{2}}\right)=\frac{\psi\left(\lvert\bm{r}_{ba_{2}}\rvert/h_{a_{2}}\right)}{\sqrt{2^{q_{1}+q_{2}}}}H_{q_{1}
}\left(\frac{x_{1,b}-x_{1,a_{2}}}{h_{a_{2}}\sqrt{2}}\right)H_{q_{2}}\left(\frac{x_{2,b}-x_{2,a_{2}}}{h_{a_{2}}\sqrt{2}}\right),\label{eq:hermite}\end{equation}
\hlb{in which $H_{q_{1}}$ is the $q_{1}$-th order univariate Hermite polynomial (of the physicists kind) and $\psi$ is an RBF, as defined in~\mbox{\cite{king_2020,king_2022}}}. Note, that whilst in~\eqref{eq:general_do} the basis functions are centred on the node at which the operator is based (i.e. $\chi$ in this case), here, the basis functions are centred on the location of node $a_{2}$. 
The consistency matrix is constructed as
\begin{equation}\bm{M}_{\chi}=\displaystyle\sum_{b\in\mathcal{N}_{\chi}}\bm{X}\left(\bm{r}_{ba_{2}}\right)\otimes\bm{W}\left(\bm{r}_{ba_{2}}\right)\end{equation}
where $\bm{X}\left(\bm{r}_{ba_{2}}\right)$ is the vector of Taylor monomials, again centred on the node $a_{2}$, and with the zeroth order monomial $1$ included. The weights $\bm{\Psi}_{\chi}^{\zeta}$ are obtained by solving
\begin{equation}\bm{M}_{\chi}\bm{\Psi}_{\chi}^{\zeta}=\bm{C}_{\chi}^{\zeta}\end{equation}
in which 
\begin{equation}\bm{C}_{\chi}^{\zeta}=\bm{X}\left(\bm{r}_{\zeta}-\bm{r}_{a_{2}}\right).\end{equation}
The resulting operator~\eqref{eq:bdo} approximates $\phi_{\zeta}$ \hlb{for $\zeta=a_{3},a_{4}$}, through an interpolation based around the location of node $a_{2}$. To ensure the computational stencil adequately samples the basis functions, the stencils for $\chi$ are extended to include all nodes which fall within the stencil of node $a_{2}$. For both $\chi=a_{0}$ and $\chi=a_{1}$, I construct $L_{\chi}^{\zeta}$ using terms up to order $m=4$ order, resulting in an interpolation which has polynomial consistency of order $m=4$ (i.e. the errors in $\phi_{\zeta}$ scale with $s_{\chi}^{5}$). 

Next I introduce the finite difference operator for some derivative of $\phi$ evaluated at node $\chi$. This operator has the same form as~\eqref{eq:general_do}, and is written as
\begin{equation}L_{\chi}^{fd}=\displaystyle\sum_{b=a_{0}}^{a_{4}}\phi_{b\chi}w_{b\chi}^{fd}.\label{eq:fdo}\end{equation}
For example, if I wish~\eqref{eq:fdo} to approximate the normal derivative on the boundary ($\chi=a_{0}$), the weights would be $w_{b\chi}^{fd}=\left[0,4,-3,4/3,1/4\right]/s_{\chi}$ for $b=\left[a_{0},a_{1},a_{2},a_{3},a_{4}\right]$. The expressions for $\phi_{\zeta}$ obtained from~\eqref{eq:bdo} can be substituted into the finite difference operator~\eqref{eq:fdo} to give 
an expression for $L_{\chi}^{fd}$ in terms of real nodes only:
\begin{equation}L_{\chi}^{fd}=\displaystyle\sum_{b\in\mathcal{N}_{\chi}}\phi_{b\chi}\hat{w}^{fd}_{b\chi}\label{eq:fdo2},\end{equation}
where I have defined 
\begin{equation}\hat{w}^{fd}_{b\chi}=\begin{cases}w_{a_{0}\chi}^{fd}+\displaystyle\sum_{\zeta=a_{3}}^{a_{4}}w_{\zeta\chi}^{fd}w_{b\chi}^{\zeta}&b=a_{0}\\w_{a_{1}\chi}^{fd}+\displaystyle\sum_{\zeta=a_{3}}^{a_{4}}w_{\zeta\chi}^{fd}w_{b\chi}^{\zeta}&b=a_{1}\\w_{a_{2}\chi}^{fd}+\displaystyle\sum_{\zeta=a_{3}}^{a_{4}}w_{\zeta\chi}^{fd}w_{b\chi}^{\zeta}&b=a_{2}\\\displaystyle\sum_{\zeta=a_{3}}^{a_{4}}w_{\zeta\chi}^{fd}w_{b\chi}^{\zeta}&\text{otherwise}.\end{cases}\end{equation}
Equation~\eqref{eq:fdo2} has the same form as~\eqref{eq:general_do}, allowing for ease of implementation of the spatial derivative operators across all internal and boundary nodes. The $5^{th}$ order consistency of the interpolants is sufficient to avoid degrading the order of accuracy of the finite difference operators on the $5$ point stencil. Although the node distribution is structured only for two rows in addition to the nodes on the boundary, the order of convergence of the derivative operators is unchanged relative to the formulation requiring the complete $5$ rows of structured nodes presented in~\cite{king_2022}. \hlb{Note that by combining the finite difference weights $w_{b\chi}^{fd}$ and the interpolation weights $w_{b\chi}^{\zeta}
$, the explicit evaluation of properties at the fictitious nodes $\zeta=a_{3}$ and $\zeta=a_{4}$ is avoided: the interpolated finite difference operators given by~\mbox{\eqref{eq:fdo2}} require only the values of $\phi$ at regular nodes.}

I note that I have also investigated using only one row of structured nodes in addition to the boundary nodes, having all other nodes unstructured, and interpolating to obtain values at a fictitious node $a_{2}$, in addition to $a_{3}$ and $a_{4}$. This approach yields derivative operators of the required polynomial consistency, and studies confirmed the order of accuracy is unchanged. However, the reduction of the region of structured nodes at the boundary increases the sensitivity of the system to node disorder: there are more situations where the solution of the consistency correction matrices on the boundary gives rise to non-negligible stagnation errors. In these situations, the stability of the derivative operators is reduced, and a much smaller time-step is required for a stable simulation. It is possible that increases in stencil size would alleviate this problem. A smaller layer of structured nodes at the boundary will provide greater flexibility in the geometries which can be discretised at a given resolution. Ultimately, the ability to discretise the domain with unstructured nodes right up to the boundary, whilst retaining a boundary discretisation with $4^{th}$ order consistency, would be a significant development, and one on which I am actively working.

\bibliography{jrckbib}

\end{document}